\documentclass[printer]{aa} % for a referee version
\usepackage{lineno}
\usepackage{graphicx}
\usepackage{epstopdf}
\usepackage{txfonts}
\usepackage{multirow}
\usepackage{longtable}

\def\fdeg{\hbox{$^\circ$}}

\begin{document}
%\linenumbers
%
\title{On the distance to the North Polar Spur and the local CO-H$_{2}$ factor%
\thanks{Based on observations obtained with XMM-Newton, an ESA science mission with instruments and contributions directly funded by ESA Member States and NASA.}
\thanks{Based on data obtained using the t\'elescope Bernard Lyot at Observatoire du Pic du Midi, CNRS and Universit\'e Paul Sabatier, France.}}
\subtitle{}

   \author{R. Lallement \inst{1}
   	\and
          S. Snowden \inst{2}
          \and
         K.D. Kuntz \inst{3}
          \and
         T.M. Dame \inst{4}
  \and
         D. Koutroumpa \inst{5}
         \and
         I. Grenier \inst{6}
         \and
         J.M. Casandjian \inst{6}}

   \institute{ 1 - GEPI/ Observatoire de Paris, 5 Place Jules Janssen, 92195 Meudon, France \email{rosine.lallement@obspm.fr} \\
2 - NASA/GSFC, Greenbelt, MD 20771, USA\\
3 - Henry A. Rowland Department of Physics \& Astronomy, Baltimore, MD21218, USA\\
4 - Harvard-Smithsonian Center for Astrophysics, 60 Garden Street, Cambridge, MA 02138, USA\\
5 - LATMOS-IPSL, Universit\'e de Versailles Saint Quentin, INSU/CNRS, 11 Bd D'Alembert, 78280 Guyancourt, France\\
6 - AIM, Université Paris Diderot \& CEA Saclay DSM/Irfu/SAp, 91191 Gif/Yvette, France}

   \date{Received ; revised }

%4 - Universit\'e de Savoie, LGIT-Savoie, 73376 Le-Bourget-du-Lac, France\\
% \abstract{}{}{}{}{}
% 5 {} token are mandatory

  \abstract
   {}
   {Most models identify the X-ray bright North Polar Spur (NPS) with a hot interstellar (IS) bubble in the Sco-Cen star-forming region at $\simeq$130 pc. An opposite view considers the NPS as a distant structure associated with Galactic nuclear outflows. Constraints on the NPS distance can be obtained by comparing the foreground IS gas column inferred from X-ray absorption to the distribution of gas and dust along the line of sight. Absorbing columns towards shadowing molecular clouds simultaneously constrain the CO-H$_{2}$ conversion factor.}
%There are also strong implications for all emission models, in particular polarized microwave Galactic foreground. We revisit here the question of the NPS location.}
   {We derived the columns of X-ray absorbing matter N$_{Habs}$ from spectral fitting of dedicated XMM-Newton observations towards the NPS southern terminus (l$^{II}\simeq29\fdeg$, b$^{II}\simeq+5$ to $+11\fdeg$). The distribution of the IS matter was obtained from absorption lines in stellar spectra, 3D dust maps and emission data, including high spatial resolution CO measurements recorded for this purpose.}
   {N$_{Habs}$ varies from $\simeq$ 4.3 to $\simeq$ 1.3 x 10$^{21}$ cm$^{-2}$ along the 19 fields. Relationships between X-ray brightness, absorbing
column and hardness ratio demonstrate a brightness decrease with latitude governed by increasing absorption. The comparison with absorption data, local and large-scale dust maps rules out a NPS near side closer than 300 pc. The correlation between N$_{Habs}$ and the reddening increases with the sightline length from 300 pc to 4 kpc and is the tightest with Planck $\tau_{353GHz}$-based reddening, suggesting a much larger distance. N(H)/E(B-V)$_{\tau}$ $\simeq$ 4.1 x 10$^{21}$ cm$^{-2}$ mag$^{-1}$, close to Fermi-Planck determinations. N$_{Habs}$ absolute values are compatible with HI-CO clouds at -5 $\leq$ V$_{LSR}$ $\leq$ +25 to +45 km s$^{-1}$ and a NPS potentially far beyond the Local Arm.
 A  shadow cast by a b=+9$\fdeg$ molecular cloud constrains X$_{CO}$ in that direction to $\leq$ 1.0 x 10$^{20}$ cm$^{-2}$ K$^{-1}$ km$^{-1}$ s. The average X$_{CO}$ over the fields is $\leq$ 0.75 x 10$^{20}$ cm$^{-2}$ K$^{-1}$ km$^{-1}$ s.}
  % conclusions heading (optional), leave it empty if necessary % co\"{i}ncidentally 
 {}
 % {The NPS source is definitely not a nearby structure, is very likely beyond the Local Arm, and potentially farther away. The nearby molecular clouds in front of the NPS have a very low X$_{CO}$ factor, in agreement with recent Fermi-Planck determinations.}
  %{The southern terminus of the NPS is fully absorption bounded and its source lies behind the Local Arm, at least three hundreds parsecs away. There is additional evidence for a much larger distance from both IS gas and dust absorption and emission data.}


 \maketitle
%
%________________________________________________________________

\section{Introduction}

The North Polar Spur is one of the best known features in radio continuum and diffuse soft X-ray background maps. It is seen as a $\simeq$ $15^\circ$ wide arc that runs with varying intensity from $l,b\sim25^\circ,20^\circ$ to $330^\circ,75^\circ$ \citep[e.g.,][]{Brown60,Bowyer68}.  As surveys of the diffuse background expanded it was seen to be one of the most prominent features of the entire 
sky, although it was joined by a wide region of diffuse emission in the general direction of the Galactic center both above and below the Galactic 
plane, the {\it X-ray bulge} \citep{Snowden95, Snowden97}.  Radio Loop I \citep{Berkhuijsen71,Haslam82} lies next to 
the NPS, and appears to bound the NPS. Both are also seen in polarized radio emission and in total-intensity and polarized
microwave emission \citep{Sofue79, Sun14, Vidal14, PlanckXXV15}.
Filaments and arcs seen in HI and extending up to +85\fdeg northern galactic latitude are also spatially associated with the Loop I and the NPS \citep{Colomb80,Kalberla05}. The NPS/Loop I is detected at GeV energies \citep{Casandjian09, Ackermann14}, presumably due to inverse-Compton scattering of starlight by the
energetic electrons, combined with pion decay emission from the
cold border.

Using Loop I to outline a small 
circle on the sky it was then assumed that the NPS was 
the limb-brightened edge of a superbubble with a radius of $\simeq100$pc 
centered on the Sco-Cen OB association at $\sim130$~pc, with the Sco-Cen 
OB association easily creating and powering the superbubble with both 
stellar winds and supernovae \citep{deGeus92,Egger95}. The size and high latitude extent of the Loop I strongly suggests its proximity, and indeed measured distances to the HI arcs, either from stellar light polarization \citep{Heiles00}, or from absorption studies \citep{Puspi12} are on the order of 100 pc. The HI shells are thought to be shock-compressed ISM at the periphery of the Loop I/NPS expanding structure. Sophisticated models of time-dependent evolution of the ISM under the action of winds and supernovae are able to reproduce the Loop I structure and its interaction with the cavity surrounding the Sun (the Local Bubble, or LB), and most of the observations \citep{deAvillez05, Breitschwerdt06}.

A different  interpretation of the NPS enhancement has been defended over the years \citep{Sofue74,Sofue94,Sofue00},  based on several arguments on the geometry and difficulties in adjusting SNR models to the radio continuum, X-ray and HI measurements.  
According to the author, NPS/Loop I better traces a shock front propagating through the Galactic halo, having originated from an intense explosion and/or a starburst at the galactic center, of energy 3 x 10$^{56}$ ergs and about 15 million years ago. While the shock can mimic the radio and X-ray North Polar Spur, the post-shock high-temperature gas may also explain the observed X-ray bulge around the Galactic center. 
%According to this scenario, the NPS radio continuum is thermal, instead of synchrotron-type, which corresponds to the NPS radio spectral properties.  
%xxx argument de la vitesse de loop I seulement moins de 20 kms pas assez si supernova pour expliquer les X 
More recently, \cite{BlandHawthorn03} uncovered a  200 pc wide bipolar structure at the Galactic Center at mid-infrared wavelengths, likely associated with a bursting episode. Interestingly, they also showed that a large scale structure extrapolated from the central region and extending up to the halo could be seen from the Sun as a Loop despite being open-ended, provided it is wider than the solar galactocentric radius, and suggested that the \cite{Snowden97} X-ray bulge observed at 0.5-2.0 keV reveals this bipolar structure.
 Finally, the existence of Galactic nuclear activity and associated large-scale structures has been spectacularly demonstrated with the discovery of the gigantic $\gamma$-ray bubbles in the high energy {\it Fermi}-LAT data \citep{Su10}. The Fermi bubbles (FBs) and concentric structures at their feet in the Galactic Plane are also seen in total-intensity or polarized radio and microwave emission \citep{Carretti13}, their inner parts are close to the X-ray bulge (see Fig. 6 from \cite{Casandjian15F}) and their edges seem to parallel bright arcs in the 1.5-2 keV ROSAT maps (see Fig.  20 from \cite{Su10}).
 
   Following the discovery several models have been proposed for the FBs: one class of scenarios considers a recent, short, AGN-type outburst activity  while other models consider less energetic, long duration Galactic wind models maintained by supernovae. In the former (latter) case the $\gamma$-ray emission is mainly of leptonic (hadronic) origin (see, e.g., \cite{Crocker15} and \cite{Sarkar15} for further discussion and description of their analytical and numerical  Galactic wind models).
 %Jones et al (2012) also found structures that look similar in polarized microwave and gamma rays, and identified them as bubble sub-structures, also located at the galactic center. 
Outside the bubbles, \cite{Su10} also identify larger gamma-ray structures, the so-called {\it inner arc}  that seems to border the low latitude portion of the northern bubble at galactic longitude +20$\fdeg$, and the {\it outer arc} at +25-;+30$\fdeg$. Both are seen in polarized microwave emission, and the {\it inner arc} is clearly considered as a galactic center feature. The {\it outer arc}, which is very similar to the inner arc, seems to parallel the low latitude part of the NPS (see Fig. 2 of  \citep{Su10}). 
%Indeed, the NPS and Loop I are also remarkably parallel to the edge of the northern polarized radio lobe or northern bubble edge. 

Based on these geometrical arguments, \cite{Su10} considered the possibility that Loop I and the northern outer arc are parts of the relics of previous bubbles.  On the other hand, while there are other X-ray arcs in the vicinity of the Galactic center, there are no structures similar to the NPS either in the northern Galactic hemisphere on the other side of the FB or at all in the south.  This would require a rather asymmetric origin mechanism.
 
The implications of the two scenarios in terms of Galactic nuclear activity, its temporal variability and outflow interaction with the halo are drastically different, and today the NPS distance and the link between NPS/Loop I and the Galactic Center are still a matter of debate.  \cite{PlanckXXV15} extensively discuss the NPS-Loop I characteristics and origin based on Planck-WMAP polarization, and list the various constraints on their distance. As noted by the authors, new evidence for a distance greater than the traditional value of 130 pc has been published. \cite{Wolleben07} showed that the radio emission from Loop I is strongly depolarized below 30$\fdeg$ latitude and attributed this depolarization to fluctuations
in the foreground Faraday depth. The required path length beyond the Local Bubble is above 70 pc, which places the front face of the loop beyond
 the Sco-Cen association. More recently \cite{Puspitarini14} failed to identify in 3D maps of the nearby IS dust \citep{Lallement14} a large cavity that could potentially be filled with hot gas and produce the NPS emission, contrary to other X-ray enhancements that have cavity counterparts. Instead, they suggested that the near side of the NPS source is located beyond 200 pc. \cite{Sofue15} used the X-ray absorption pattern and inferred that the source is beyond the Aquila Rift main dense part, i.e. again beyond the Sco-Cen association (but see section 2). On the other hand, \cite{PlanckXXV15} concluded based on geometrical arguments and the Planck polarization maps that the NPS is not associated with the Galactic Center. 

Here we address the question of the NPS distance by means of a detailed comparison between the columns of interstellar gas that are located in front of the NPS source and absorb its X-ray emission, and the distribution with distance  of the IS matter derived from stellar absorption data. The former information is obtained by means of spectral fitting of new, dedicated X-ray observations, and the latter from dedicated ground-based spectroscopic data and published dust maps. We complement this study by  additional comparisons between X-ray absorption and IS emission data. Dedicated CO measurements at high spatial resolution were performed for this purpose.

In section 2 we describe the new XMM-Newton data and their spectral analysis. In section 3 we present the ground-based optical data and compare them as well  as 3D reddening maps with the X-ray absorbing columns deduced from the XMM fitting results. In section 4 we present the new CO data and compare X-ray absorbing columns with line-of-sight IS dust and gas based on emission data. In section 5 we summarize the comparisons and draw conclusions.

\section{X-ray observations of the NPS southern terminus}

\subsection{RASS Data}
\begin{figure}
\centering
\includegraphics[width=0.98\linewidth,angle=0]{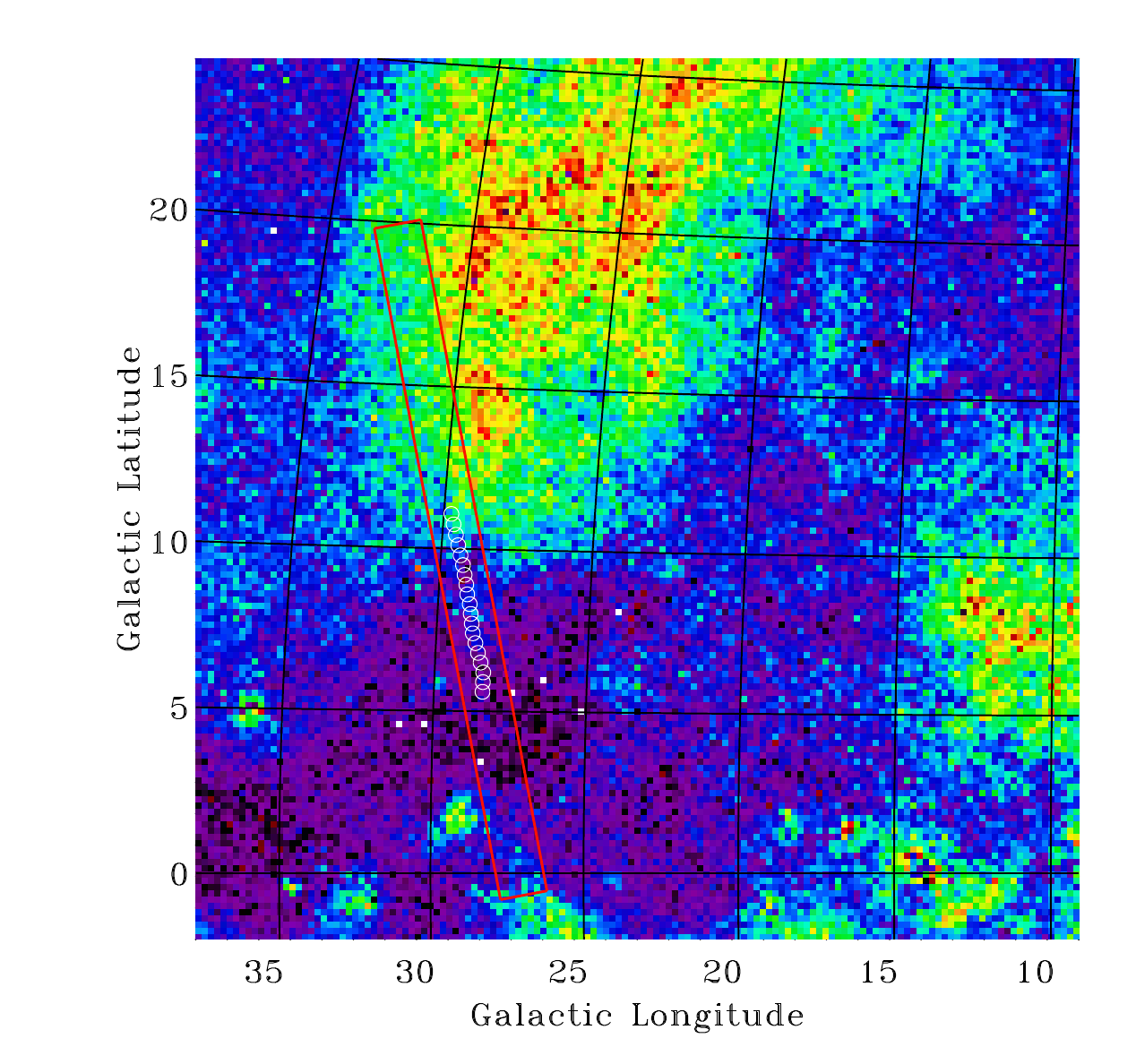}
\caption{The RASS R4+R5+R6 band image of the NPS. The color scale is linear and runs from 0 (black) to $1.5 x 10^{-3}$ count s$^{-1}$ arcmin$^{-2}$ (red). The white circles show the location of the {\it XMM-Newton} pointings. The red box is the region from which was extracted the {\it ROSAT} profile shown in the next figure.}
\label{sofue_imag}
\end{figure}

\begin{figure}
\centering
\includegraphics[width=0.98\linewidth,angle=0]{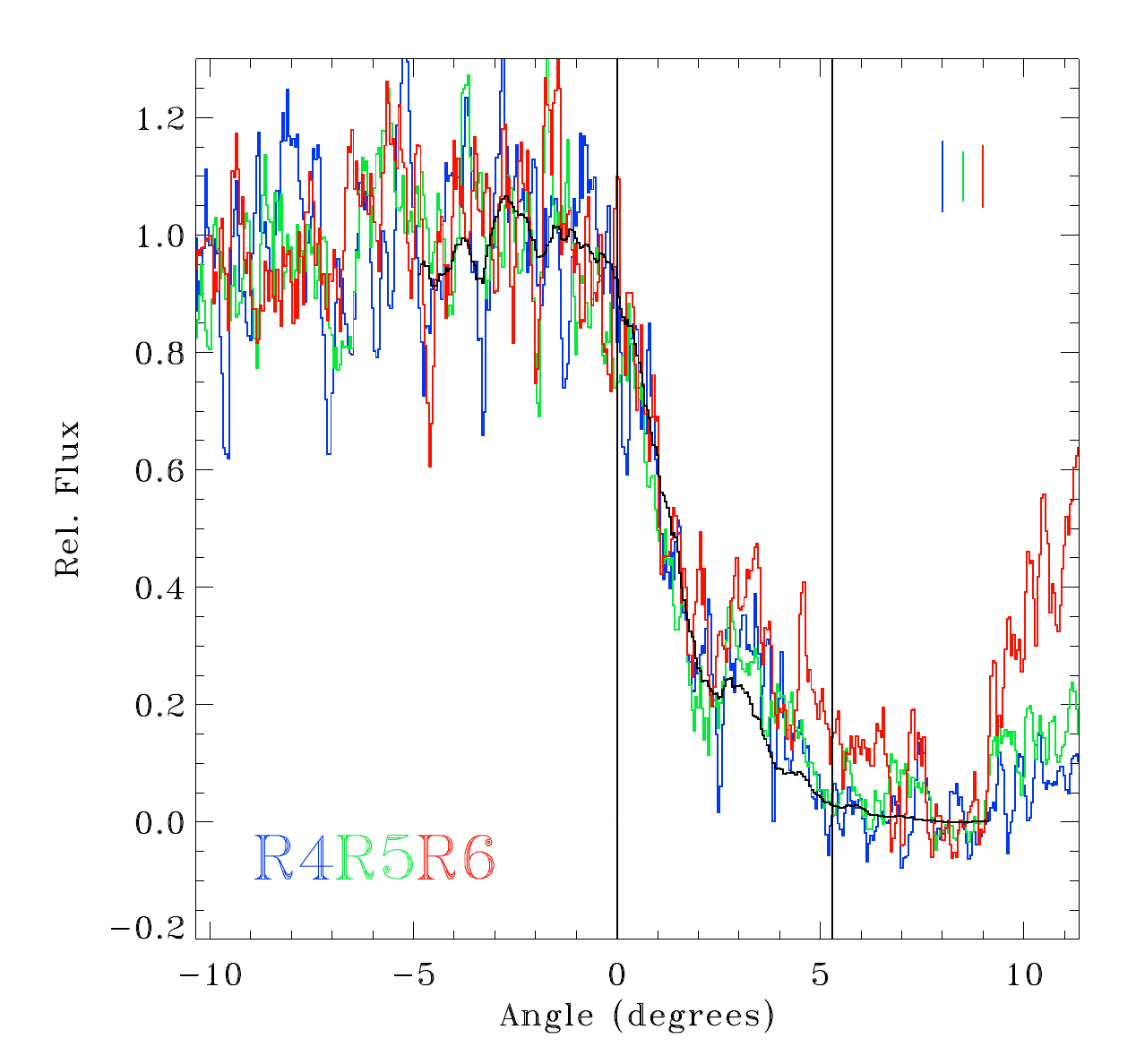}
\caption{The profile of the emission in the RASS R4, R5, and R6 bands along the box shown in the previous figure.
In each case the profile has had the minimum emission level removed, and has been normalized so that the emission in the brightest part of the profile (the NPS) is set to unity. The black histogram shows the expected absorption profile.
%This is simply the best fit A exp(-IRAS/B).
The perpendiculars show the extent over which we obtained the {\it XMM-Newton} data. The bars in the upper right show the typical uncertainty of the {\it ROSAT} data in each band where the absorption is changing most strongly.}
\label{sofue_prof}
\end{figure}

Near the Galactic plane the NPS has an angular width of roughly ten degrees, and at all latitudes it has a low characteristic temperature (on the order of 0.2 keV \citep{Willingale03,Miller08}), hence producing very little emission above 1.5 keV.  Thus {\it ROSAT}, with its 2$\fdeg$ diameter FOV, large grasp, and strong soft response, and the {\it ROSAT} All-Sky Survey (RASS) would seem to be the ideal tools with which to study the structure of the NPS. In what follows we demonstrate that it is not the case, and that more spectrally resolved data and higher signal are necessary to draw useful conclusions. Fig. \ref{sofue_imag} shows a RASS map of the Southern end of the NPS, the location of our {\it XMM-Newton} observations, and the location of a more extensive region over which we began a {\it ROSAT} study. Fig. \ref{sofue_prof} shows the profiles of the X-ray emission in the {\it ROSAT} R4 (0.44-1.01 keV), R5 (0.56-1.21 keV), and R6 (0.73-1.56 keV) bands. From the profile in each band we have subtracted that level of emission seen in the most highly absorbed part of the Galactic plane (N(H)$>10^{22}$ cm$^{-2}$) and normalized the peak of the NPS emission to unity. There are a number of reasons why this process is na\"{i}ve, but it demonstrates that within the uncertainties all three bands show essentially the same absorption profile. If the NPS were absorption bounded, one would expect the profile of the higher energy bands to extend to lower latitudes than the profiles of the lower energy bands (as discussed by \cite{Sofue15}). Fig. \ref{sofue_prof} shows that the energy resolution and the count-rate limited angular resolution of the RASS, $12\arcmin$, does not allow one to definitely diagnose whether the edge of the NPS is absorption or emission bounded. Similarly the uncertainties in the {\it ROSAT} data do not allow one to determine the absorbing column as a function of position.

\subsection{XMM-Newton Data}
\begin{figure}
\centering
\includegraphics[width=0.98\linewidth,angle=0]{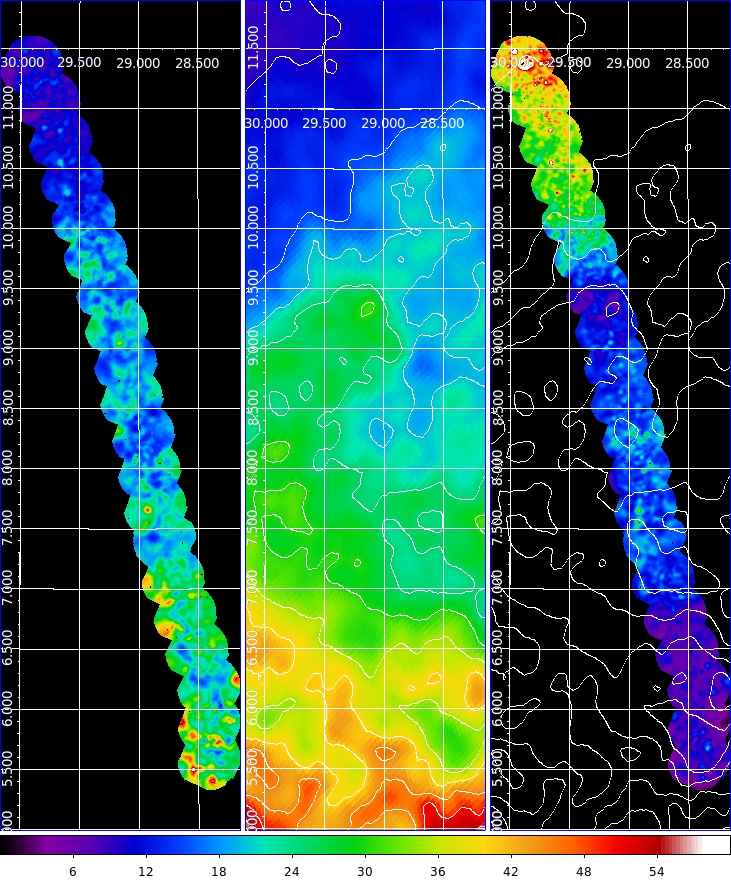}
\caption{The 19 NPS pointings: shown are the (0.75-1.3)/(0.4-0.75) keV hardness ratios (left),  IRAS 100$\mu$  data (middle) and 0.3-1.4 keV brightnesses (right). The color scale applies to the three quantities: from 0 to 4 for the hardness ratio, from 0 to 65 MJsr$^{-1}$ for IRAS and from 0 to 6 cts s$^{-1}$ deg$^{-2}$ for the 0.3-1.4 keV brightness.}
\label{fields1}
\end{figure}

A series of pointed observations with XMM-Newton of the southern terminus of the NPS were proposed to determine whether it is
emission or absorption bounded (GO program P074189, P.I. K. Kuntz) and to use the comparison between the absorption and the Galactic distribution of
cooler, X-ray absorbing material to determine the NPS source location. The positions of the pointings were chosen to: 1) cover a range of 3/4 keV surface brightnesses covering the apparent southern terminus of the NPS, 2) lie near the transverse center of the NPS, 3) include a serendipitous XMM pointing at the northern end (shown in Fig 1 but finally not used in the analysis due to a lower integration time and the loss of the center part due to the necessary exclusion of the point source) and avoid a serendipitous pointing with a bright point source at the southern end, and 4) sample the apparent absorption feature near b$\simeq$9$\fdeg$ (see Fig \ref{fields1}).
19 partially overlapping 0\fdeg.5 diameter fields provide full coverage from b=5.6 to 11\fdeg.1 at an average longitude l= 29\fdeg.  Images of the surface brightness and spectrum
hardness ratio are displayed in Fig \ref{fields1} along with the IRAS 100 $\mu$m data. There is a global decrease of the X-ray brightness with decreasing latitude, and an associated hardening of the spectra. The dust distribution is not monotonously varying with latitude, instead there is an IR maximum at b$\simeq$ 9$\fdeg$, and an associated X-ray darkening clearly visible in Fig \ref{fields1} (right panel). As we will see in section 4, it corresponds to a molecular cloud visible in the CO maps.

The processing of the XMM data (background modeling and subtraction, point source excision, exposure correction) followed the procedures outlined in \cite{Snowden08} based on the calibration of \cite{Kuntz08}.  The XMM Extended Source Analysis Software\footnote{ftp://legacy.gsfc.nasa.gov/xmm/software/xmm-esas/xmm-esas-v13.pdf} procedures are incorporated in the XMM Science Analysis System software\footnote{http://www.cosmos.esa.int/web/xmm-newton/what-is-sas}.

\begin{figure}
\centering
\includegraphics[width=0.95\linewidth,angle=0]{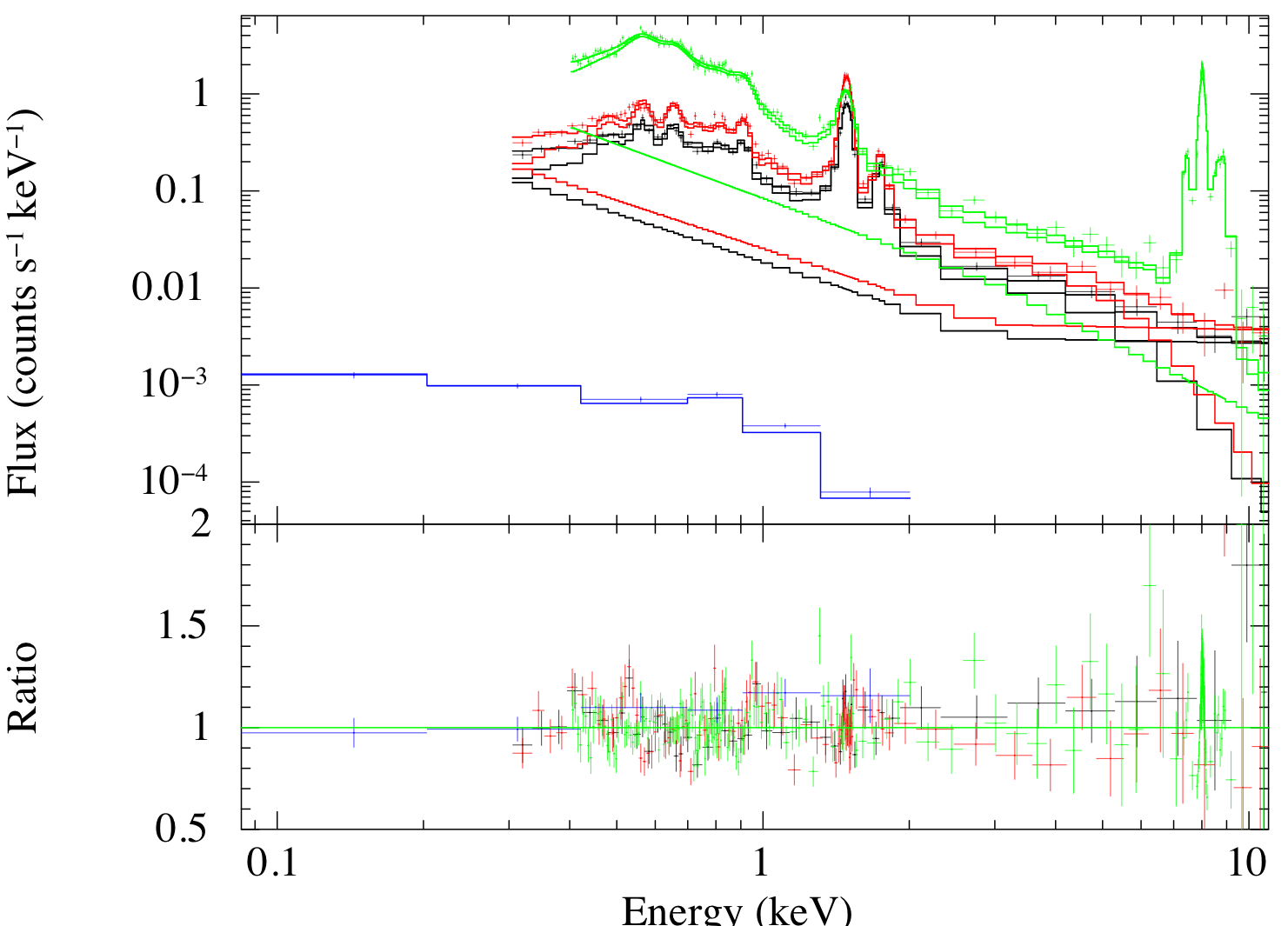}
\caption{Fitted MOS1 (black), MOS2 (red), and pn (green) spectra for the NPS pointing number 17 along with the RASS spectrum (blue) for a 1 degree diameter region centered on the pointing.  For each of the EPIC spectra there are three curves that show the fitted broken power law representing the soft proton background (lower for all), the fitted cosmic spectrum plus fluorescent background lines (middle for all), and the the sum of the two representing the best fit to the data.  The MOS fluorescent lines are the Al K alpha and Si K Alpha at 1.49 and 1.74 keV, respectively.  The pn fluorescent lines are the Al K alpha and a variety of Cu, Ni, and Zn lines around 8 keV.}
\label{fits1}
\end{figure}

\begin{figure}
\centering
\includegraphics[width=0.98\linewidth,angle=0]{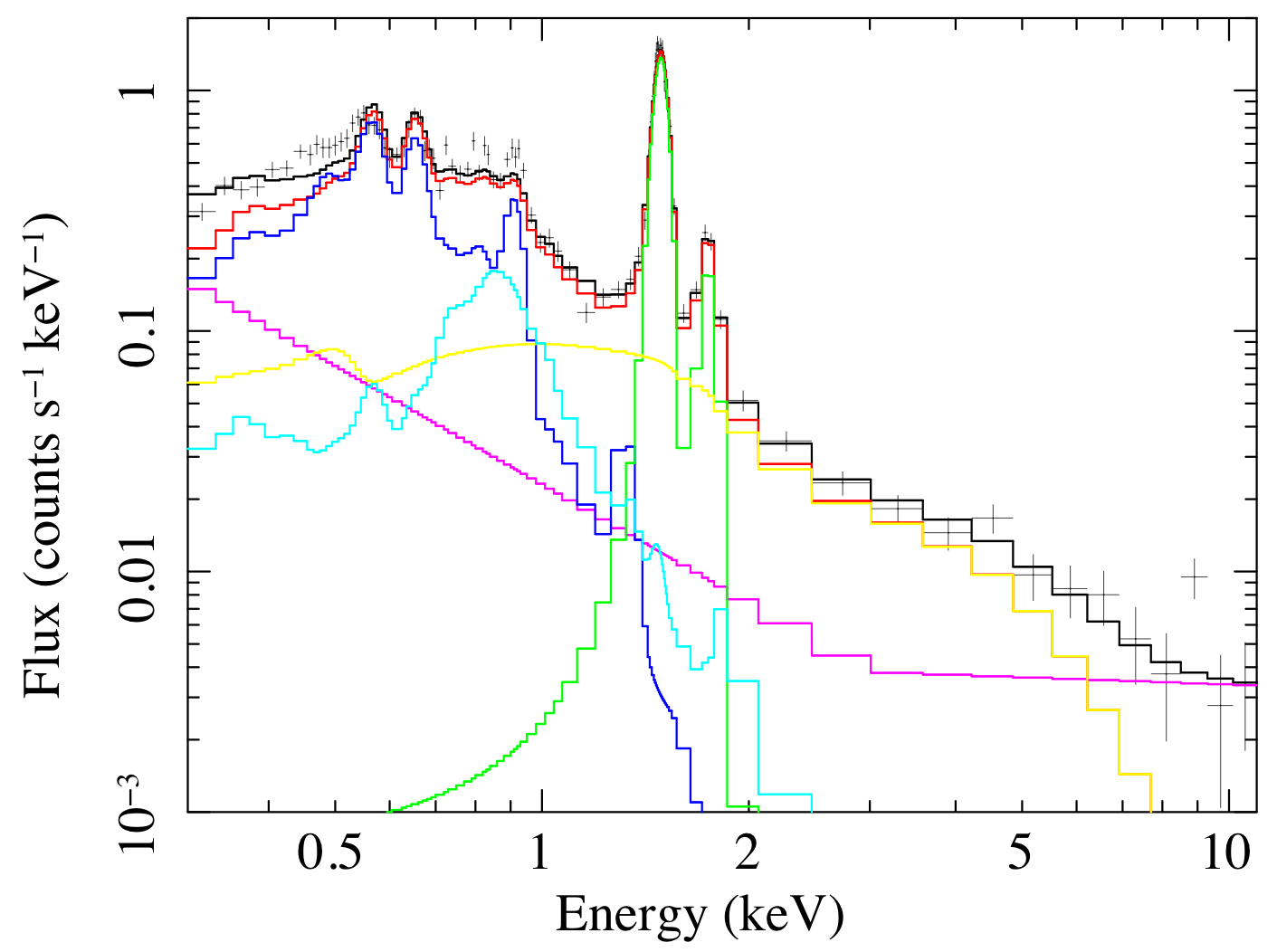}
\caption{Fitted MOS2 spectrum from the NPS \#17 pointing showing the individual components: Dark Blue - NPS, Light Blue - Galactic Background, Yellow - Extragalactic Background, Green - Instrumental Fluorescent Background, Red - Sum of the Above, Violet - Soft Proton Background, Black Curve - Sum of All Components}
\label{fits2}
\end{figure}

\begin{table*}
\caption{Observed XMM fields and fit results for freely varying NPS and Galactic Bulge fluxes and absorbing column densities.}
\begin{center}
\begin{tiny}
\begin{tabular}{|c|c|c|c|c|c|c|c|c|}
\hline 
EXP & Field & Gal. long  & Gal. lat   & N${H}$(NPS)  & F(NPS) unabsorbed& F(NPS) transmitted & N${H}$(bulge) & F(bulge) \\
 &  & ($^\circ$)  &  ($^\circ$)  &  10$^{20}$ cm$^{-2}$ & 10$^{-7}$ ergs s$^{-1}$ cm$^{-2}$ sr$^{-1}$    &  10$^{-8}$ ergs s$^{-1}$ cm$^{-2}$ sr$^{-1}$  & 10$^{20}$ cm $^{-2}$ & 10$^{-7}$ ergs s$^{-1}$ cm$^{-2}$ sr$^{-1}$   \\
\hline 
0760080101&01a&28.4&5.6&40.0$\pm$4.5&1.99$\pm$1.21 &0.75$\pm$0.48 &90.6$\pm$4.9 &1.35$\pm$0.24\\
0741890201&02&28.4&5.9&30.5$\pm$7.6&0.67$\pm$0.75 &0.43$\pm$0.16 &57.7$\pm$10.3&0.46$\pm$0.31\\
0741890301&03&28.4&6.2&36.6$\pm$3.6&1.73$\pm$0.88 &0.79$\pm$0.40 &88.9$\pm$5.1 &1.03$\pm$0.23\\
0741890401&04&28.5&6.5&41.4$\pm$4.3&2.14$\pm$1.30 &0.77$\pm$0.46&101.1$\pm$8.0 &1.40$\pm$0.41\\
0741890501&05&28.6&6.8&46.8$\pm$3.7&4.14$\pm$2.13 &1.04$\pm$0.53 &49.9$\pm$12.9&0.31$\pm$0.23\\
0741890601&06&28.7&7.1&35.7$\pm$3.0&3.35$\pm$1.48 &1.60$\pm$0.71 &63.8$\pm$4.9 &1.10$\pm$0.26\\
0741892201&07a&28.8&7.4&18.2$\pm$3.4&1.74$\pm$0.88 &2.59$\pm$1.32 &62.3$\pm$6.8 &1.25$\pm$0.43\\
0741890801&08&28.85&7.7&35.8$\pm$3.5&2.78$\pm$1.41 &1.32$\pm$0.67 &59.4$\pm$5.5 &0.88$\pm$0.25\\
0741890901&09&28.9&8&26.3$\pm$3.1&1.96$\pm$0.93 &1.65$\pm$0.79 &47.4$\pm$6.7 &0.63$\pm$0.26\\
0741891001&10&28.95&8.3&21.8$\pm$1.9&2.34$\pm$0.69 &2.64$\pm$0.78 &65.4$\pm$3.6 &0.87$\pm$0.18\\
0741891101&11&29.05&8.6&29.8$\pm$3.2&1.89$\pm$0.98 &1.28$\pm$0.66 &38.3$\pm$8.1 &0.36$\pm$0.21\\
0741891201&12&29.1&8.9&28.0$\pm$2.9&1.86$\pm$0.85 &1.40$\pm$0.64 &38.9$\pm$9.8 &0.31$\pm$0.24\\
0741891301&13&29.175&9.2&36.1$\pm$2.7&2.96$\pm$1.22 &1.38$\pm$0.57 &48.1$\pm$11.2&0.33$\pm$0.28\\
0741891401&14&29.25&9.5&18.4$\pm$3.0&0.97$\pm$0.46 &1.39$\pm$0.65 &50.0$\pm$12.2&0.37$\pm$0.33\\
0760080201&15a&29.35&9.8&24.9$\pm$2.4&2.37$\pm$0.88 &2.17$\pm$0.81 &28.5$\pm$7.6 &0.30$\pm$0.18\\
0760080301&16a&29.45&10.11&21.5$\pm$2.4&2.40$\pm$0.84 &2.77$\pm$0.97  &0.0$\pm$2.3 &0.12$\pm$0.03\\
0741891701&17&29.55&10.43&20.4$\pm$1.8&3.67$\pm$0.99 &4.53$\pm$1.23  &0.0$\pm$2.5 &0.14$\pm$0.03\\
0741891801&18&29.65&10.75&16.9$\pm$1.8&2.98$\pm$0.82 &4.70$\pm$1.29  &0.0$\pm$2.9 &0.15$\pm$0.04\\
0741891901&19&29.75&11.07&14.7$\pm$1.5&2.93$\pm$0.69 &5.41$\pm$1.28  &0.4$\pm$0.1 &0.26$\pm$0.02\\
\hline
\end{tabular}
\end{tiny}
\end{center}
\label{tabfields}
\end{table*} 

\begin{table*}
\caption{Fit results for a unique NPS flux over the $\simeq$ 5$\deg$ latitude interval, a flux linearly varying with latitude with the average slope deduced from the free-flux case, and a slope reduced by a factor of two. Fit qualities are very similar for the three cases (see text).}
\begin{center}
\begin{tiny}
\begin{tabular}{|c|c|c|c|c|}
\hline 
EXP & Field & N${H}$(NPS) cst & N${H}$(NPS) lin x0.5 & N${H}$(NPS) lin\\
 &  &  10$^{20}$ cm $^{-2}$ & 10$^{20}$ cm $^{-2}$ &10$^{20}$ cm $^{-2}$\\
\hline 
0760080101&01a&42.7$\pm$1.3 &38.5$\pm$1.3&35.3$\pm$1.4 \\
0741890201&02 &50.1$\pm$1.6 &46$\pm$1.6&43.0$\pm$1.6 \\
0741890301&03 &41.4$\pm$1.3 &38.1$\pm$1.3&35.5$\pm$1.3 \\
0741890401&04 &45.2$\pm$1.3 &39.5$\pm$1.3&37.2$\pm$1.3 \\
0741890501&05 &38.8$\pm$1.3 &36.4$\pm$1.4&34.6$\pm$1.4 \\
0741890601&06 &30.5$\pm$1.2 &28.5$\pm$1.2&26.9$\pm$1.2 \\
0741892201&07a&22.3$\pm$1.2 &20.5$\pm$1.2&19.0$\pm$1.3 \\
0741890801&08 &33.3$\pm$1.2 &31.7$\pm$1.2&30.5$\pm$1.3 \\
0741890901&09 &28.7$\pm$1.2 &27.4$\pm$1.2&26.4$\pm$1.2 \\
0741891001&10 &21.8$\pm$1.1 &20.8$\pm$1.2&20.0$\pm$1.2 \\
0741891101&11 &32.5$\pm$1.2 &31.7$\pm$1.3&31.1$\pm$1.3 \\
0741891201&12 &31.2$\pm$1.2 &30.7$\pm$1.3&30.2$\pm$1.3 \\
0741891301&13 &32.6$\pm$1.2 &32.4$\pm$1.3&32.1$\pm$1.3 \\
0741891401&14 &31.4$\pm$1.2 &31.5$\pm$1.2&31.2$\pm$1.2 \\
0760080201&15a&24.5$\pm$1.2 &24.8$\pm$1.2&24.8$\pm$1.3 \\
0760080301&16a&21.0$\pm$1.2 &21.6$\pm$1.3&21.7$\pm$1.3 \\
0741891701&17 &14.2$\pm$1.2 &15.1$\pm$1.4&15.3$\pm$1.3 \\
0741891801&18 &13.6$\pm$1.2 &14.6$\pm$1.3&14.9$\pm$1.3 \\
0741891901&19 &11.6$\pm$1.1 &12.8$\pm$1.2&13.3$\pm$1.2 \\
\hline
\end{tabular}
\end{tiny}
\end{center}
\label{tabfields2}
\end{table*}

\begin{figure}
\centering
\includegraphics[width=0.98\linewidth,height=8cm, angle=0]{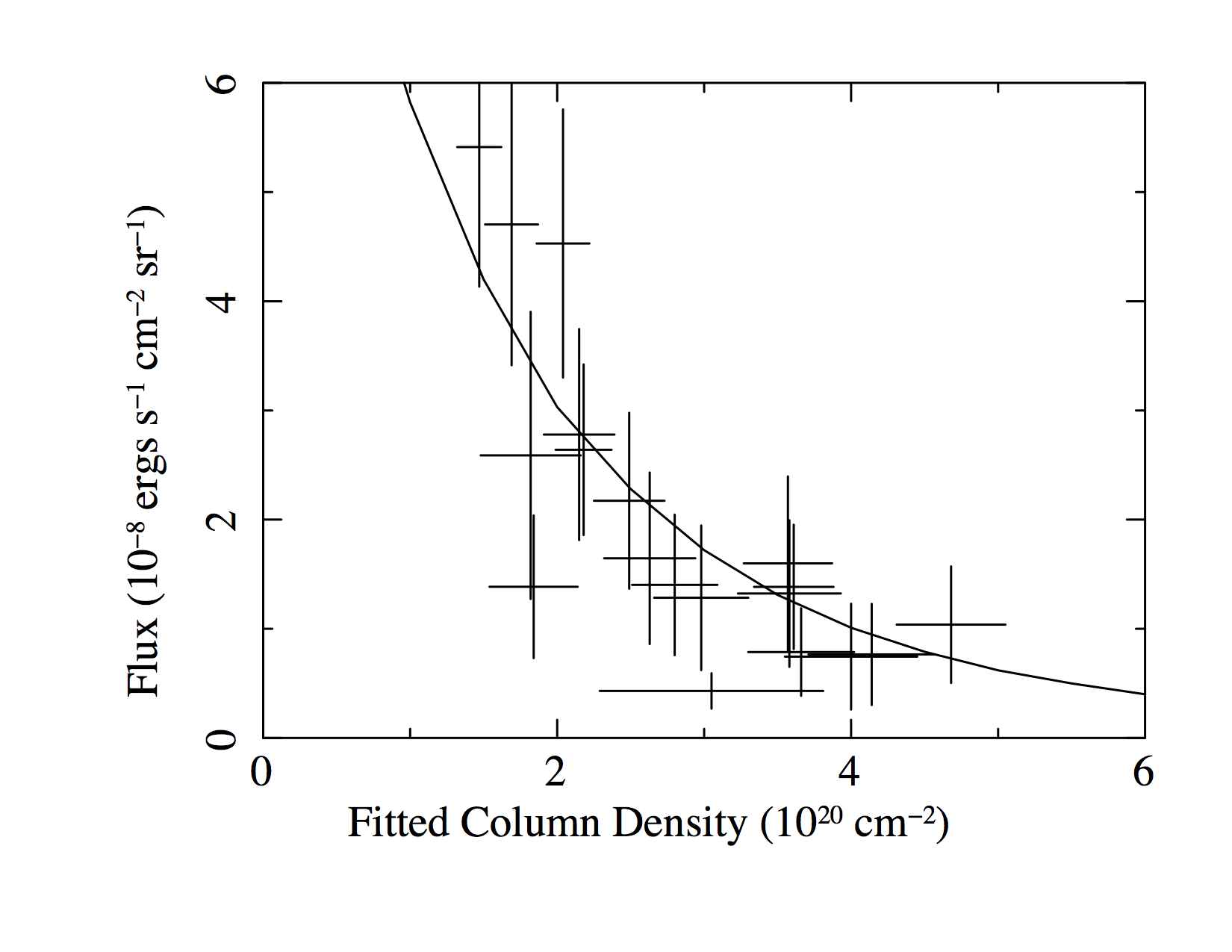}
\caption{Fitted NPS measured brightness as a function of the fitted absorbing column, in the free flux case. The curve is the model X-ray band-average absorption roughly scaled to the data.}
\label{FLUX_ABSORB}
\end{figure}

\begin{figure}
\centering
\includegraphics[width=0.9\linewidth, angle=0]{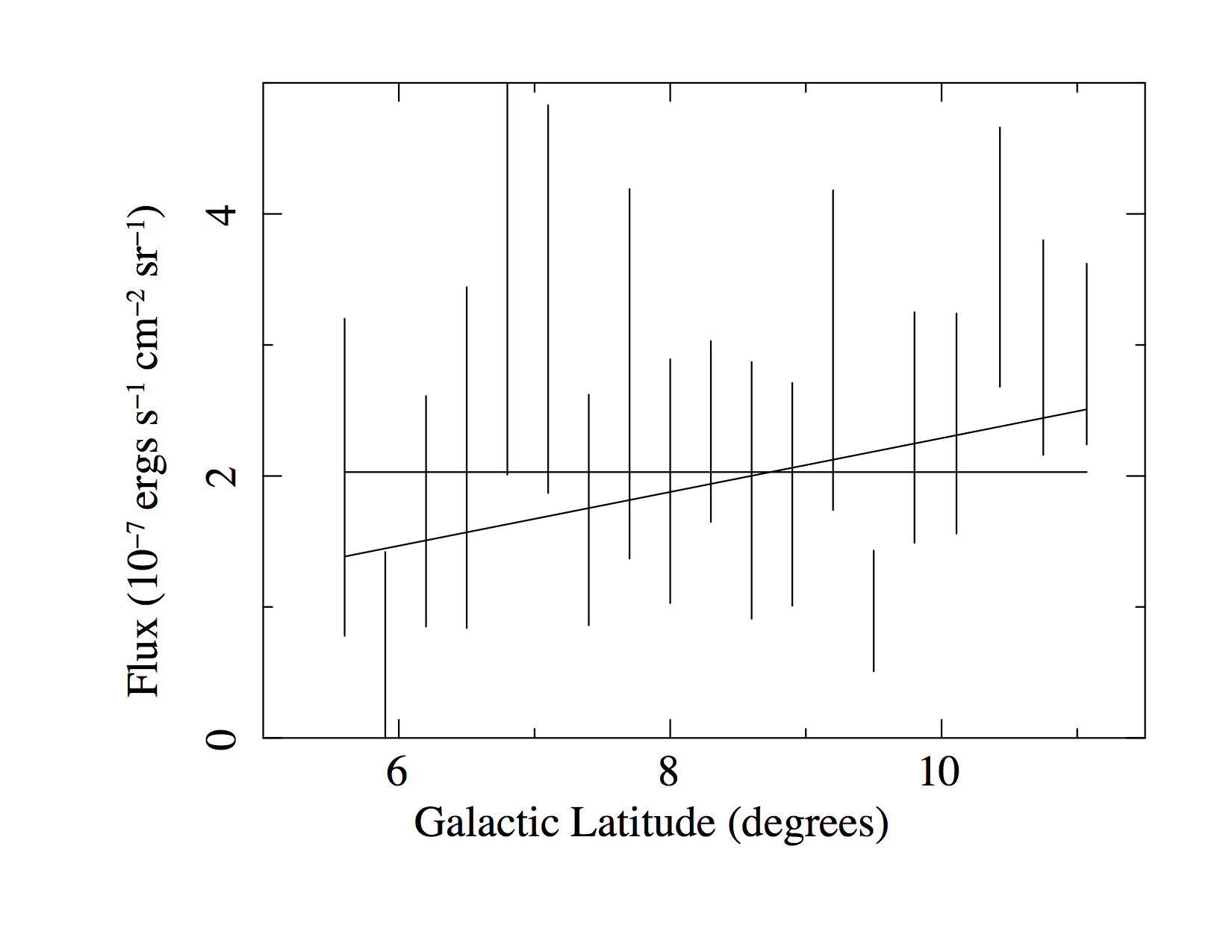}
\caption{Fitted NPS fluxes for the 19 pointings.}
\label{figfluxes}
\end{figure}

All MOS1, MOS2 and pn spectra were fitted simultaneously for the 19 fields, i.e. 57 spectra, along with three RASS spectra to constrain the Local Hot Bubble (LHB) contribution (see Fig \ref{fits1} and \ref{fits2}. The models include three thermal emission components (APEC models with \cite{Anders89} abundances), (i) an unabsorbed LHB component characterized by a single low temperature, (ii) an absorbed hotter component representing the NPS, and (iii) a hot Galactic bulge component. A cosmic X-ray background contribution (CXB) in the form of a power-law with a spectral index of 1.46 is also included in the model. The contribution from heliospheric solar wind charge exchange is assumed to be small and included in the LHB contribution.

The first step in the X-ray spectral analysis process was to fit the RASS spectra separately to determine the LHB emission temperature and flux (kT$_{LHB}$=0.111$\pm$0.006 keV, F$_{LHB}$= 1.33$\pm$0.06 x 10$^{-8}$ ergs s$^{-1}$ cm$^{-2}$ sr$^{-1}$).  These parameters were then held fixed in all subsequent fits.  Next, the 57 XMM spectra and 3 RASS spectra were simultaneously fit with the temperature for the NPS emission tied to the same value for all spectra and the temperature for the Bulge emission likewise.  The fluxes for both the NPS and the Bulge, along with their associated absorption column densities, were allowed to vary individually for all pointings.  The fitted temperatures are T$_{NPS}$=0.179$\pm$0.002 keV and T$_{Bulge}$= 0.702$\pm$0.007 keV, and the fitted fluxes and absorption column densities are listed in Table \ref{tabfields}.  The fit had a $\chi^{2}$ value of 25974 for 17779 degrees of freedom.  The fitted NPS absorbing column densities and transmitted intensities are displayed in Fig \ref{FLUX_ABSORB}, along with an appropriate absorption curve (determined by absorbing the NPS thermal spectrum using WebPIMMS\footnote{http://heasarc.gsfc.nasa.gov/cgi-bin/Tools/w3pimms/w3pimms.pl}). Although there is considerable scatter, the data are reasonably consistent with the distribution expected by absorption of a relatively uniform distant emission component by foreground material for all observation directions.  This demonstrates that the NPS brightness decrease toward lower latitudes is primarily governed by the foreground absorption precluding an emission-bounded terminus, and strongly favors a NPS source located  farther toward the Galactic plane. 

With the large number of both spectra and parameters the spectral fitting process was slow, and made more complicated by the strong correlation between many parameters such as the flux and absorption column density.  However, these preliminary results show that the NPS emission continues toward the Galactic plane as far as the observations extend.  Fig. \ref{figfluxes} shows the fitted values for the NPS flux as a function of Galactic latitude, along with both constant and linear fits to the data.  For the third stage of the spectral analysis we fitted  1) a tied constant value to the NPS flux ($\chi^2=26031.3$), 2) flux values tied with the linear fit results in Fig. 6 ($\chi^2=26024.3$), and 3) values tied with the slope reduced by a factor of two ($\chi^2$=26024.6). In all three cases the tied flux values were allowed to float and there were 17797 degrees of freedom. The fitted values for the ISM column densities for the three cases are listed in Table 2. The fitted values are similar and show the same trends.  Differences between the resulting columns give an estimate of the systematic uncertainties associated with the fitting method. Until section 4.2 in this paper we will use the results from case number 3. 

 To summarize the XMM spectral results, the best fit values for the LHB, NPS and Bulge parameters are: kT$_{LHB}$ =  0.111 keV, kT$_{NPS}$ =  0.181 keV, kT$_{Bulge}$ = 0.705 keV, F$_{LHB}$ = 1.34 x 10$^{-8}$ ergs s$^{-1}$  cm$^{-2}$  sr$^{-1}$, F$_{NPS}$ increases linearly between 1.45 x 10$^{-7}$ and 2.62 x 10$^{-7}$ ergs s$^{-1}$  cm$^{-2}$  sr$^{-1}$ with increasing Galactic latitude, and F$_{Bulge}$ generally decreases from 1.33 x 10$^{-7}$ to 0.16 x 10$^{-7}$ ergs s$^{-1}$  cm$^{-2}$  sr$^{-1}$ with increasing Galactic latitude.  The fitted values for the NPS foreground N$_{Habs}$ vary from 4.3 to 1.3 x 10$^{21}$ cm$^{-2}$ along the 19 fields generally decreasing with increasing Galactic latitude and the Bulge N$_{Habs}$ generally decreases from 9.7 to 0.0 x 10$^{21}$ cm$^{-2}$ (the latter value giving an idea of some of the systematics in the fits). The fitted NPS column is equal or above the columns found from spectral analysis of Suzaku and XMM NPS data by Miller et al. (2008) and Willingale et al. (2003) at higher latitude and similar longitude (NH(abs) =  0.7 x 10$^{21}$ cm$^{-2}$  at (l,b)=(25,+20), 0.5 x 10$^{21}$ cm$^{-2}$ at (l,b)=(27,+22), 0.33 x 10$^{21}$ cm$^{-2}$ at (l,b)=(20,+30), and 0.2 x 10$^{21}$ cm$^{-2}$ at (l,b)=(20,+40).

\section{Comparison between X-ray absorbing columns and $\leq$ 1 kpc absorption measurements}

\subsection{High-resolution NARVAL optical spectra and interstellar KI profile-fitting}

\begin{figure}
\centering
\includegraphics[width=0.9\linewidth,angle=0]{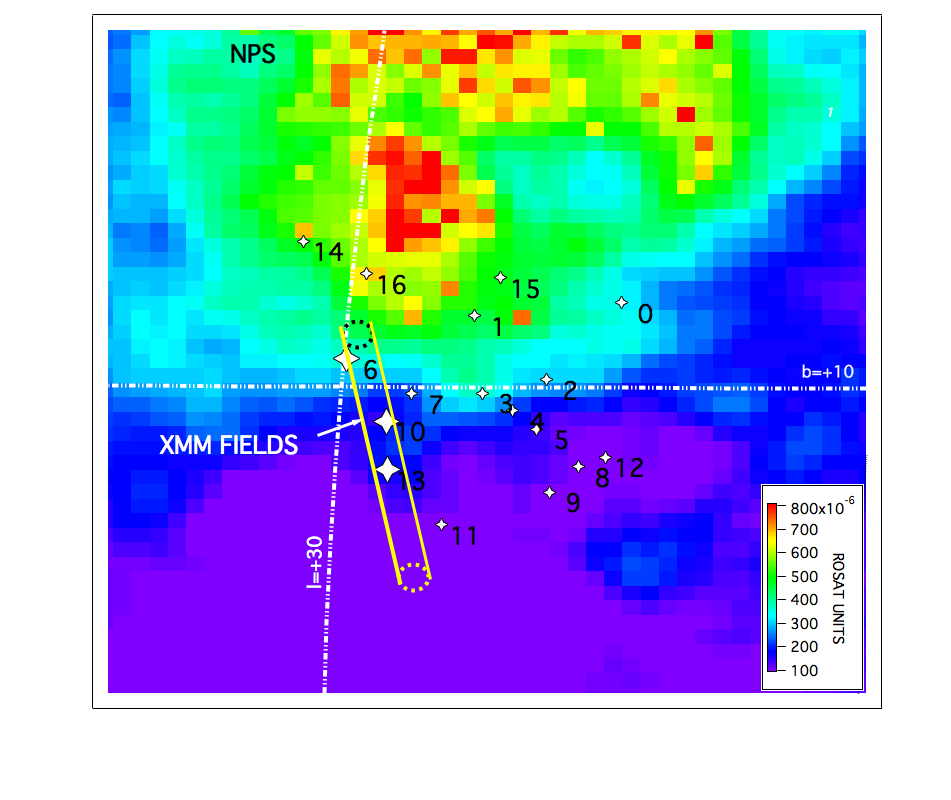}
\caption{NPS XMM pointing and ground-based targets (white diamonds), superimposed on the ROSAT 0.75 keV map. The three targets that fall along the XMM path are shown as larger signs.}
\label{fields3}
\end{figure}

We have selected  a series of nearby target stars located in the direction of the NPS southern terminus and possessing a Hipparcos parallax distance.
The targets stars are shown in Fig \ref{fields3} superimposed on the RASS 3/4 keV map.  The targets are early-type stars or in an opposite way cool G-K stars having a radial velocity different from the mean interstellar motion in this area.   High signal to noise ratios and high resolution (R=80,000) spectra were recorded with NARVAL, the  spectro-polarimeter of the Bernard Lyot telescope (2m) at Pic du Midi Observatory, used in the spectrometric mode. Observations were distributed during the summer of 2014 in the frame of a dedicated program (P.I. R. Lallement). We used the standard reduction pipeline. Telluric lines were removed using our {\it rope length} method described in \cite{Raimond12} and using TAPAS synthetic transmittance spectra \citep{TAPAS14}, except for the strongest lines around the 7665 \AA\ KI line where we kept a mask in regions corresponding to the deepest telluric absorptions. In the case of the early-type stars we fitted the spectra simultaneously for the 5889–5895 \AA\ NaI-D doublet and the two 7665-7699 \AA\  KI lines, allowing for multiple clouds and polynomial functions for the stellar continua and using classical Voigt profiles (see, e.g., \cite{Welsh10}. The combined use of NaI and KI allowed us to better constrain the cloud parameters and constituted a validity check for the KI absorptions in the region contaminated by the strong telluric lines (see Fig. 19 in the Appendix). For three targets that were strongly contaminated we fitted only the KI 7698 \AA\ transition (Fig. 20). Late-type stars absorption lines were treated differently depending on the difference between the stellar and IS  radial velocities. For large differences the stellar line was simply treated as a continuum (Fig. 21). For small differences the stellar radial velocities were determined preliminarily from the whole spectrum, and simulated KI stellar lines were added to the fitting model at the star velocity (Fig. 22). In all cases a good estimate of the KI column density could be determined. The KI columns were converted into equivalent N(H) using the empirical formula established by \cite{Welty01} based on $\simeq$ 50 high-resolution spectra (formula 3 in their Table 2). The target characteristics, Hipparcos data, fitted KI columns and equivalent N(H) columns are listed in Table \ref{tabstars}.

\begin{table*}
\caption{Observed stars, their Hipparcos numbers and distances, interstellar KI absorbing columns and estimated line-of-sight gas column and color excess. Also listed is the color excess obtained from integration within the 3D local map. Stars along the XMM path are indicated by a (*) sign.}
\begin{center}
\begin{tiny}
\begin{tabular}{|c|c|c|c|c|c|c|c|c|c|c|c|c|c|c|c|}
\hline \#&HIP&glon&glat&sptype&dist&distmin&distmax&N(KI)&N(H)&E(B-V)&(+)&(-)&ebv(3D)&(+)&(-)\\
  && ($^\circ$)  &  ($^\circ$)  & & pc &pc &pc&  cm $^{-2}$ & cm $^{-2}$    & mag         &   & & mag & &  \\
\hline
0&87633&24.82&11.98&A0&183&169&199&3.26e+11&1.33e+21&0.19&0.07&0.11&0.20&0.03&0.03\\
1&88148&27.59&11.66&A3&135&126&146&1.34e+11&8.11e+20&0.10&0.04&0.07&0.06&0.01&0.01\\
2&88376&26.12&10.18&B9&227&197&269&8.42e+11&2.26e+21&0.34&0.12&0.19&0.20&0.05&0.04\\
3&88671&27.31&9.86&G8II&215&195&240&2.3e+11&1.10e+21&0.15&0.06&0.09&0.15&0.03&0.03\\
4&88698&26.72&9.46&A3&235&175&355&3.18e+11&1.31e+21&0.18&0.07&0.11&0.19&0.08&0.07\\
5&88753&26.24&9.03&B9&549&369&1075&9.4e+11&2.41e+21&0.36&0.13&0.20&0.31&0.04&0.07\\
6*&88841&29.95&10.65&A2&135&127&144&4e+08&3.15e+19&0.00&0.03&0.03&0.06&0.01&0.01\\
7&88870&28.65&9.85&K2&366&286&510&1.46e+11&8.51e+20&0.11&0.05&0.07&0.20&0.03&0.03\\
8&88878&25.41&8.16&G5&244&213&285&1.9e+11&9.86e+20&0.13&0.05&0.08&0.23&0.06&0.04\\
9&89148&25.93&7.57&K0&204&175&244&2.8e+11&1.22e+21&0.17&0.07&0.10&0.13&0.03&0.09\\
10*&89151&29.10&9.21&A2&279&204&444&4e+10&4.13e+20&0.04&0.02&0.03&0.16&0.05&0.05\\
11&89680&27.92&6.80&A0&228&192&279&1.04e+12&2.55e+21&0.39&0.14&0.21&0.17&0.08&0.05\\
12&88740&24.91&8.38&F8&95&90&100&0&0&0.00&0.03&0.03&0.03&0.01&0.01\\
13*&89472&29.00&8.08&K2&344&257&518&5.28e+11&1.74e+21&0.25&0.10&0.15&0.20&0.04&0.05\\
14&88149&31.00&13.37&B2V&200&190&211&2.8e+10&3.38e+20&0.02&0.02&0.03&0.12&0.01&0.01\\
15&87812&27.16&12.55&B2IV-V&417&356&503&1.36e+12&2.96e+21&0.45&0.16&0.25&0.28&0.01&0.01\\
16&88192&29.72&12.63&B5I&377&313&474&2.3e+11&1.10e+21&0.15&0.06&0.09&0.19&0.01&0.01\\
\hline
\end{tabular}
\end{tiny}
\end{center}
\label{tabstars}
\end{table*}

\subsection{First constraints from the target stars along the XMM path}
Three of the target stars are located within our XMM fields, hip88841(star 6), hip89151(star 10), hip89472 (star 13), shown in Fig \ref{fields3}. This allows a comparison between their distance-limited N(H) columns deduced from KI absorption lines and the measured XMM N$_{Habs}$ (see Table \ref{tabstars}). For the three targets the optical columns are either significantly below or slightly below the XMM absorbing columns, strongly suggesting that the NPS source is located beyond the stars. Their Hipparcos distances are 135(-8,+9), 279(-65,+175) and  344(-87,+175) pc resp., i.e. the largest of the three minimal distances is 257 pc. We conclude that the near side of the NPS source is more than 250 pc distant. 

\subsection{3D map of nearby dust and absorption measurements}

\begin{figure}
\centering
\includegraphics[width=0.7\linewidth,angle=0]{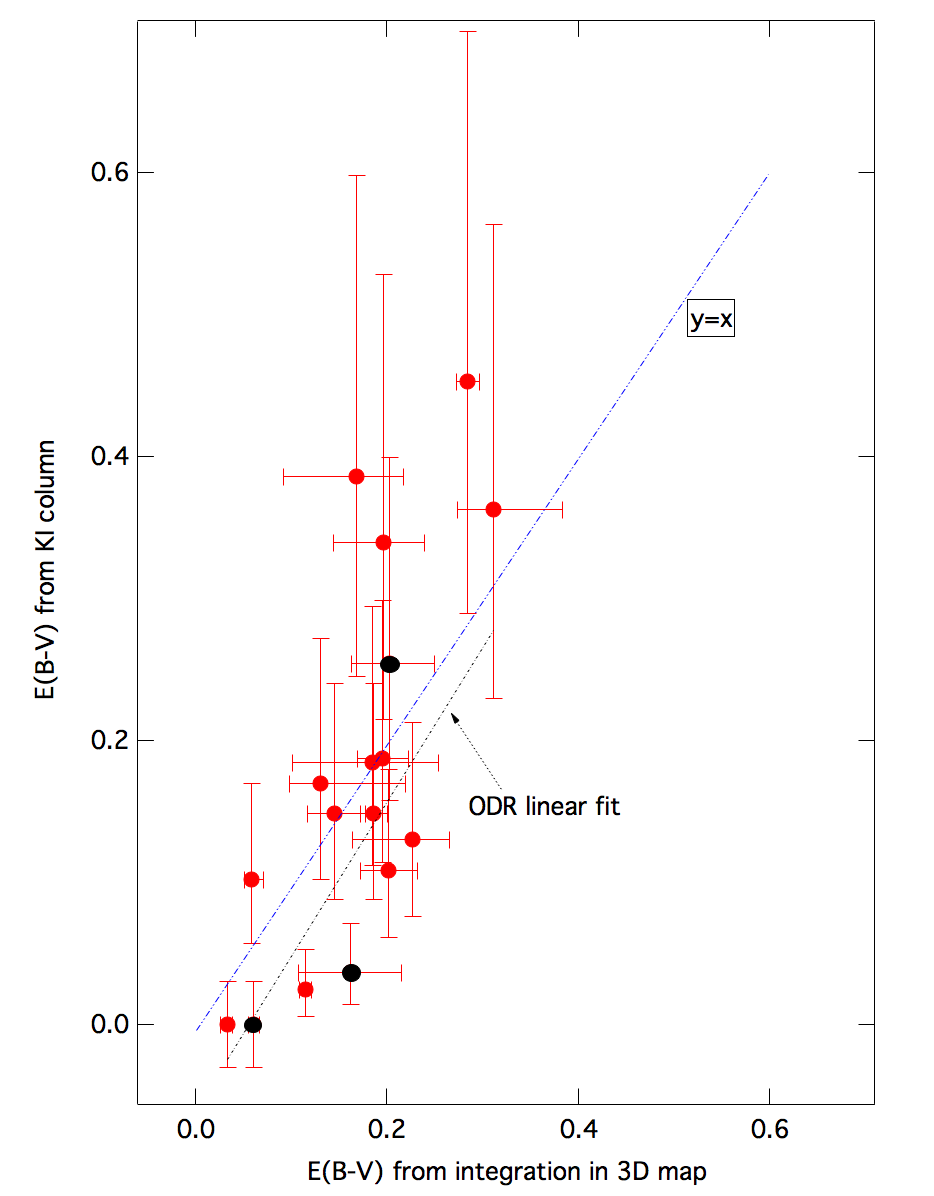}
\caption{Estimated color excess of the target stars based on their KI lines, using the empirical relationship of \cite{Gudennavar12}, and estimated reddening through integration within the local dust map. A linear fit taking into account errors on both quantities 
(orthogonal distance regression, ODR)  is superimposed. The three targets stars along the XMM path are shown in black.}
\label{KI_EBV}
\end{figure}

%The series of absorption measurements allows us to locate the main clouds in the sky area corresponding to the NPS southern terminus and establish a bridge between the 3D reddening map of \cite{Lallement14} that is restricted to the nearby matter and large scale Pan-STARRS maps (see the next section) that become precise beyond a few hundreds parsecs. 
The whole Narval absorption dataset is used to validate the computed 3D map of local dust of \cite{Lallement14} in the NPS southern terminus area. The maps are obtained through inversion of individual color excess measurements and they have two limitations: first, there is a minimum size for the inverted structures, due to the limited number of target stars. Second, there is a bias towards weakly reddened stars and for this reason there are many fewer targets in the database that are located beyond opaque clouds than in front of them. Those limitations influence the distance range over which the maps can be used safely in a complex way and it is useful here to obtain an independent confirmation of the dust reddening found by inversion, especially here for the Aquila rift area. To do so we have integrated through the computed 3D map the color excess E(B-V) from the Sun to each target location, and did it separately for their Hipparcos minimal, most probable  and maximal distances. The resulting color excesses are listed in Table \ref{tabstars}. We compared these map-integrated values with the KI-based estimated H columns. Fig \ref{KI_EBV} shows the comparison and reveals a clear increasing trend but a large dispersion. Such departures from linearity are not surprising due to the very low spatial resolution of the map and also due to the variability of the H/KI ratio. The important result here is that within error bars the best linear fit is compatible with the empirical relationship between KI and E(B-V) found by \cite{Gudennavar12}, as shown in the figure. This implies that in this area, and for H columns on the order of those measured to our targets, an integration through the map provides a reasonable order of magnitude of the reddening. Since our targets have well defined distances up to $\simeq$ 300pc, this means that we do not miss significant structures up to this distance. At larger distance, however, the 3D maps become too uncertain due to the lack of reddening measurements (see \cite{Lallement14} for more information). 

Based on this validity check we have integrated through the maps from the Sun to 300 pc along the 19 XMM field directions. The integrations are shown in Fig \ref{LOS_ALL}. Given the small angular range covered by the XMM pointings, the reddening does not vary more than by about 30$\%$ from minimum to maximum reddening, and E(B-V) does not exceed 0.25, which is the color excess value reached at the lowest latitude. Such reddenings correspond to N(H) columns on the order of 1.0 x 10$^{21}$ cm$^{-2}$, well below the XMM measured absorptions at low latitudes. Also shown in Fig. \ref{LOS_ALL} are the color excess values estimated from KI for the field stars. Using both types of data it can be seen from the figure that there is a first increase in opacity at about 100-150 pc, followed by a larger increase distributed between 200 and 400 pc. This corresponds to the well-known Aquila rift clouds. The variability from star to star confirms the complex structure of those clouds, as shown in Fig. 8 of \cite{Puspitarini14}. Nevertheless, the comparison between the reddening achieved at 300 pc along the XMM path based on the local maps and the XMM absorption column densities very strongly suggests a NPS source beyond 300 pc, beyond the second {\it wall} revealed by the map and star absorptions.

\subsection{Use of local and Pan-STARRS 3D reddening maps}

\cite{Green15} have produced 3D reddening maps based on the Pan-STARRS (PS) photometric survey that possess a high degree of angular resolution and map up to 5-10 kpc depending on directions. These maps are used here to derive the reddening as a function of distance along the XMM fields.  They become precise only beyond about 300-400 pc and are ideally complementary to the local maps. To do so, we used the on-line tool built by the authors and extracted the reddening measurements and uncertainties for 29 directions separated by 0.1$\fdeg$ within each XMM circular field. We averaged only those data that have a good quality flag. At short distance, namely 400 and 600 pc, the number of valid points is small for some fields and in a few cases there is a large dispersion that is simply due to uncertainties distances of the the closest clouds, themselves linked to statistical uncertainties of the photometric method. We conservatively kept measurements in fields possessing at least 5 well clustered valid points and in all fields we excluded 2-sigma outliers. 

\begin{figure*}
\centering
\includegraphics[width=0.7\linewidth, angle=0]{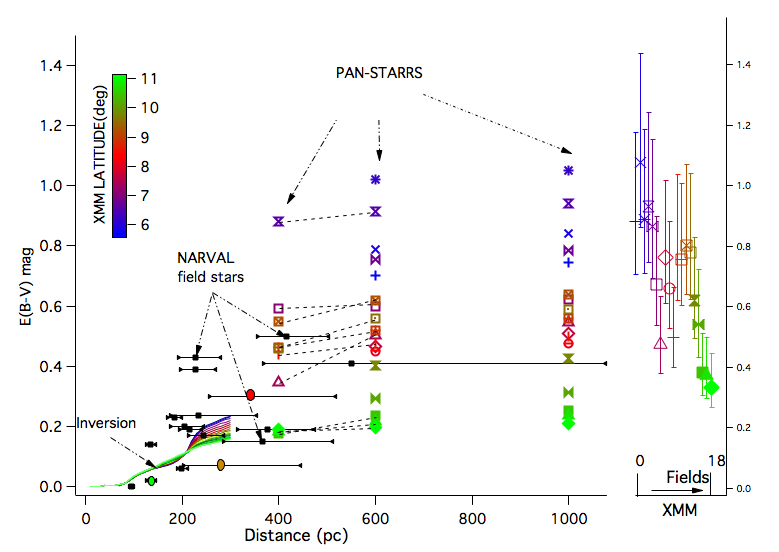}
\caption{Left panel: Color excess E(B-V) estimates for the 19 XMM field directions, colored according to the Galactic latitude: (i) integration in the local reddening inverted maps (solid lines);  (ii) Pan-STARRS color excess at 400, 600 and 1000pc from \cite{Green15}, rescaled values using measurements at 4kpc and Planck $\tau$-based E(B-V)s (see text). Colors refer to the galactic latitude (inserted color scale) and markers are different for each direction. Some measurements are missing shorter than 1kpc, due to their too large uncertainty). Also shown are the color excess values based on KI absorption lines for the NARVAL field stars in the NPS terminus region as a function of their Hipparcos distance ranges (black dots and error bars). The three targets that are located along the XMM path are indicated by large circles whose colors refer to the latitude. Right panel: XMM NPS fitted absorbing columns scaled to E(B-V) using N(H)= 3 to 5 x 10$^{21}$ cm$^{-2}$ mag$^{-1}$.}
\label{LOS_ALL}
\end{figure*}

As discussed by \cite{Green15}, in the Aquila region the PS reddening  is overestimated, potentially due to stellar metallicity gradients that are not taken into account or to peculiar properties of the dust grains.  For this reason we have chosen to scale the PS E(B-V)'s values to Planck measurements based on the 353GHz dust optical depth $\tau$353. To do so we first correlated the PS reddenings with the Planck reddenings, averaged on the same 29 directions in each field, at increasing distances. As expected, the correlation coefficient increases with distance, and at 4 kpc the Planck and Pan-STARRS E(B-V)s are nearly perfectly correlated (Pearson= 0.992). This is expected because at this distance and for the XMM field latitudes the distance to the Plane is Z$\geq$400 pc, far above the dust scale height that is on the order of 130 pc. Beyond 4 kpc the correlation between PS and Planck starts to decrease due to an observational bias. As a matter of fact, in strongly reddened fields and at large distance the number of target stars becomes insufficient for the statistical analysis and there is a favored selection of the less reddened directions, which results in an average reddening lower than the actual one. For these reasons, we considered the 4kpc measurement as the best estimate of the reddening over the entire sightline. We thus fitted the PS(4kpc) - Planck relationship to a second order polynomial and applying the derived relationship to the whole set of PS data (the mean ratio between the PS and Planck measurements is on the order of 1.25, in agreement with Fig. 12 from  \cite{Green15}). Table \ref{tabstars} lists the resulting PS values at 400, 600, 1000 pc and 4000 pc respectively and the Planck E(B-V)s.  Fig \ref{LOS_ALL} displays as a function of distance the reddening integrated through the local 3D map, the reddening estimated from the KI absorption in the field stars and these Planck-scaled PS reddening values. Although, as discussed for a long time (see, e.g., \cite{Arce99, Bonifacio00, Cambresy01},  photometric reddenings are generally smaller than than those estimated from dust emission, especially in the case of strong extinctions, 
the use of these various diagnostics traces the evolution of the reddening with distance rather well.
%This comes in addition to the fact that the SFD determination is globally higher than photospheric determinations for low latitude-dense areas, as discussed e.g. by . For those reasons we also considered the converted value: E(B-V)corr= 0.1 +0.65 (E(B-V)-0.1) (formula restricted to E(B-V) higher than 0.1).

\begin{figure*}
\centering
\includegraphics[width=0.7\linewidth, angle=0]{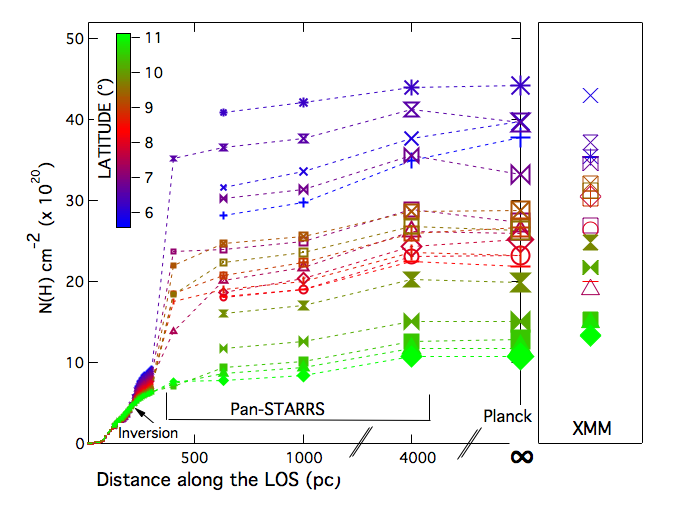}
\caption{{\bf Left panel}: From left to right for the 19 field directions, colored according to the Galactic latitude, and as a function of distance: (i) estimated gas column between the Sun and 300 pc from integration in the local reddening maps  (ii) gas columns estimated from \cite{Green15} Pan-STARRS color excess measurements up to 4 distances (iii)  gas columns estimated from Planck $\tau$-353GHz reddening measurements. See the text and Fig \ref{LOS_ALL} for the Pan-STARRS reddening scaling. All reddening are converted into N(H) using N(H)= 4. x 10$^{21}$ cm$^{-2}$ mag$^{-1}$, the central value of Fig \ref{LOS_ALL}. {\bf Right panel}:  soft X-ray absorbing column N$_{Habs}$ deduced from global fitting of XMM spectra with fixed NPS flux. The vertical scale is the same as in the left panel.}
\label{LOS_ALL_NH}
\end{figure*}

Indeed, the combination of 3D maps in Fig \ref{LOS_ALL} reveals a strong reddening jump between 300 pc and 600 pc for latitudes below +7$\fdeg$ (blue signs), with some variations in the jump location from one direction to the other. This wall of dense IS matter is also detected in the spectra of the target stars located beyond 300 pc (small black circles and distance intervals in Fig \ref{LOS_ALL}). For those directions that possess a reliable PS measurement at 400 pc, it can be seen that there is a small or significant reddening gradient  between 400 and 600 pc, as illustrated in the figure by the connecting dashed lines. This suggests that the cloud complex that produces the reddening jump extends beyond 400 pc. On the other hand, there is a plateau between 600 and 1000 pc, which allows us to conclude that the cloud complex far side is located between 400 and 600 pc and not farther out. In order to infer the NPS location with respect to this wall of dense structures, we converted the XMM absorbing gas columns into E(B-V)s using the broad interval N(H)= 4 $\pm$1 x 10$^{21}$ cm$^{-2}$ mag$^{-1}$, an interval significantly large to account for inhomogeneity of the dust to gas ratio. This interval corresponds to the most representative recent results of \cite{PlanckFermi15} who have taken into account in their analysis the contributions of the HI, DNM and CO-bright phases. Fig \ref{LOS_ALL} shows the resulting reddening values and how they compare with the previously discussed distance-limited reddening measurements or estimates. The comparison very strongly suggests that the X-ray absorbing matter corresponds to the totality of the cloud complex that is distributed between 300 and 600 pc. More specifically, the absorptions match the 600pc (or equivalently 1kpc) reddening values for a reasonable conversion factor on the order of 4. x 10$^{21}$ cm$^{-2}$ mag$^{-1}$. On the other hand, using the lower N(H)-E(B-V) factor gives reddenings marginally compatible with the 400 pc data (10 directions only). As a consequence, given the uncertainty on the conversion factor and the absence of knowledge of the cloud complex precise outer boundary distance, we very conservatively conclude that the NPS source near side lies beyond at least 300 pc. This result totally precludes any nearby source, in particular a source associated with the Sco-Cen star forming region. On the other hand, we note that the NPS near side is potentially as far as 1kpc or more, since it is not possible to determine at this stage wether the small reddening increase that is detected between 400 pc and 600 pc is seen or not in the XMM absorptions, nor the next small increase detected between 1kpc and 4kpc in some directions, see also Fig. \ref{LOS_ALL_NH}).

\begin{table*}
\caption{Dust reddening along the XMM directions: EBV$_{loc}$ is the color excess integrated from the Sun to 300 pc through the local dust map. The Pan-STARRS (PS) color excess values are averaged over the field and scaled to match Planck$_{\tau}$ at 4kpc (see text).}
\begin{center}
\begin{tiny}
\begin{tabular}{|c|c|c|c|c|c|c|c|c|c|c|c|}
\hline
Field& EBV$_{loc}$& PS-0.4kpc & PS-0.6kpc & PS-1kpc & PS-2kpc & PS-3.16kpc& PS-4kpc & Planck$_{\tau}$\\
 	&    mag & mag & mag &mag & mag & mag & mag & mag\\
\hline 
01A&0.23&&0.70&0.75&0.81&0.86&0.87&0.95\\
02&0.23&&0.79&0.84&0.91&0.94&0.94&0.99\\
03&0.23&&1.02&1.05&1.10&1.10&1.10&1.10\\
04&0.22&0.88&0.91&0.94&1.01&1.03&1.03&0.99\\
05&0.21&&0.76&0.78&0.85&0.89&0.89&0.83\\
06&0.20&0.59&0.60&0.62&0.68&0.72&0.72&0.68\\
07&0.20&0.35&0.50&0.54&0.59&0.65&0.65&0.65\\
08&0.19&&0.47&0.51&0.55&0.61&0.61&0.63\\
09&0.19&&0.45&0.48&0.53&0.58&0.58&0.58\\
10&0.18&&0.45&0.48&0.52&0.56&0.56&0.55\\
11&0.18&0.44&0.48&0.50&0.54&0.59&0.59&0.58\\
12&0.17&0.46&0.52&0.56&0.61&0.65&0.65&0.67\\
13&0.17&0.55&0.62&0.64&0.70&0.72&0.72&0.72\\
14&0.17&0.46&0.56&0.59&0.64&0.67&0.67&0.66\\
15A&0.16&&0.40&0.43&0.46&0.50&0.51&0.50\\
16A&0.16&&0.29&0.31&0.34&0.38&0.38&0.38\\
17&0.16&0.18&0.23&0.25&0.28&0.32&0.32&0.32\\
18&0.16&0.18&0.21&0.23&0.26&0.29&0.29&0.29\\
19&0.16&0.19&0.19&0.21&0.25&0.27&0.27&0.27\\
\hline
\end{tabular}
\end{tiny}
\end{center}
\label{tabxmmlos}
\end{table*} 

\begin{table*}
\caption{Correlations between the X-ray absorbing columns (Y) and color excesses. N(H)abs= B x E(B-V). (1) NPS flux varying linearly with latitude; (2) Linear variation: slope x (0.5); (3) NPS flux constant over the fields.}
\begin{center}
\begin{tiny}
\begin{tabular}{|c|c|c|c|c|c|c|c|c|c|}
\hline
 & 1 & 1 & 1 & 2& 2 & 2 & 3& 3 & 3\\
d (pc) &$\chi^{2}_{red}$& B & $\sigma$(B) &$\chi^{2}_{red}$& B & $\sigma$(B) &$\chi^{2}_{red}$& B & $\sigma$(B)\\
\hline
600&33.6&48.5&2.3&30.7&50.8&2.2&26.7&53.9&2.1\\
1000&30.2&46.3&2.1&27.1&48.5&2.0&23.1&51.5&1.9\\
2000&26.9&43.0&1.9&23.9&45.0&1.7&20.4&47.8&1.6\\
3160&22.5&41.0&1.6&19.6&42.8&1.5&16.9&45.5&1.4\\
4000&21.9&41.0&1.6&19.0&42.9&1.5&16.3&45.5&1.4\\
Planck&19.8&41.0&1.5&16.3&42.9&1.4&13.0&45.5&1.2\\
\hline
\end{tabular}
\end{tiny}
\end{center}
\label{tabcorrel}
\end{table*} 

\begin{figure}
\centering
\includegraphics[width=0.8\linewidth,angle=0]{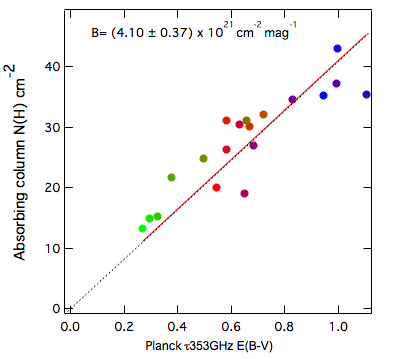}
\caption{XMM gas absorbing column (linear case) vs Planck E(B-V) $\tau_{353}$ (averaged over the XMM FOV). The color coding is for Galactic latitude and is the same as in Figs. \ref{LOS_ALL} and \ref{LOS_ALL_NH}.}
\label{xmmplanck}
\end{figure}

%\begin{figure}[h!]
%\centering
%\includegraphics[width=0.98\linewidth, angle=0]{GraphNHCO_IG_TD.png}
%\caption{Comparison between NPS X-ray absorbing gas column for the 19 fields (black diamonds) and columns from radio emission data (dashed and solid lines): shown are average (H) columns deduced from HI 21 cm EBHIS data and new CO data for various velocity intervals. The X$_{CO}$ conversion factor is assumed to be 0.5 10$^{21}$ cm$^{-2}$ K$^{-1}$ km$^{-1}$ s (see text).}
%\label{NH_WCO} 
%\end{figure}

\section{Comparison between X-ray absorbing columns and absorption over larger distances}

\subsection{Dust}

Based on the previous results we extended our comparisons to larger distances and additionally used emission data as tracers of the total amount of dust and gas along the fields. In the case of the dust, we extended our comparisons to PS at $\simeq$3.2 and 4 kpc and to Planck. For Planck we used the color excess E(B-V) map derived from the dust optical thickness $\tau$353GHz \citep{PlanckEBV14}, and averaged the reddening E(B-V)$_{\tau353}$ over the XMM fields of view in the same way as described above for the correlation with PS at 4 kpc. We averaged the PS data and scaled them in the way described  in the previous section (averaged and scaled data in Table \ref{tabxmmlos}). Fig \ref{LOS_ALL_NH} shows the reddenings from the local map, PS determinations at 400, 600, 1000 and 4000 pc and Planck, this time all converted into equivalent N(H) columns using 4.0 x 10$^{21}$ H cm$^{-2}$ mag$^{-1}$, a value appropriate for the local matter, as stated in the previous paragraph. Also plotted are XMM NPS absorbing columns, which allows a direct visual comparison between the absorptions and the reddening as a function of distance. The figure shows that, despite seemingly random discrepancies for the 5 lowest latitudes, overall the NPS absorbing columns are on the order of the converted E(B-V)s for distances as  large as 4 kpc and 
equivalently to the Planck values. However, the smallness of the reddening increase from the Aquila rift clouds to large distances , i.e., say, between 600pc and 4 kpc, and the model uncertainties on N$_{Habs}$ does not allow to definitely disentangle between these two limits.

If the NPS source lies beyond all clouds located within a given distance, and this distance is the same along the 5 $\fdeg$ XMM path, there must be a similar latitude profile between the reddening and N(H). Table \ref{tabcorrel} compares the standard deviations from proportional linear relationships relating the measured absorbing columns to the PS reddenings at varying distances and to Planck. The table shows that, for the three fitting methods, the correlation improves for increasing distances, and is optimal for the Planck reddening. Although part of the effect may be linked to the improved statistics of the PS measure between 600 pc and 3 kpc, and the quality of the Planck determination, those correlations suggest that most of the dust detected by PS and Planck lies in front of the NPS source. Fig \ref{xmmplanck} shows the correlation between N$_{Habs}$ and Planck $\tau$353GHz-based E(B-V). The slope of the proportional linear relationship is 4.0 x 10$^{21}$ H cm$^{-2}$ mag$^{-1}$, a reasonable value, al ready stated.

\begin{figure}
\centering
\includegraphics[width=0.9\linewidth,angle=0]{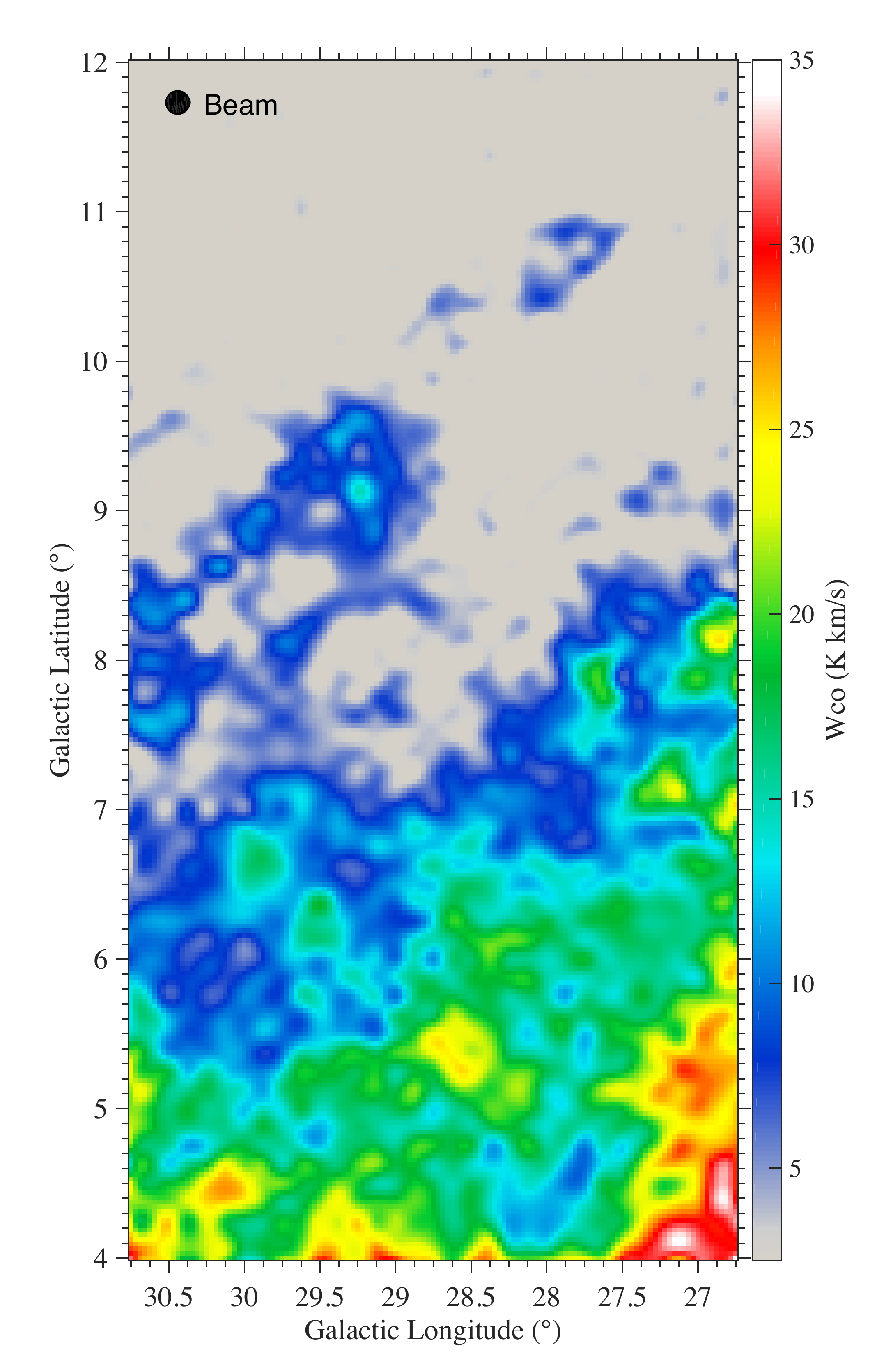}
\caption{Our new CO survey of the XMM fields, integrated from -5 to +25 km s${-1}$, the full range over which significant emission is detected.  The survey was sampled every 7.5’ with the 8.4’ beam shown at upper left. }
\label{New_CO_MAP}
\end{figure}

\begin{figure}
\centering
\includegraphics[width=0.9\linewidth,angle=0]{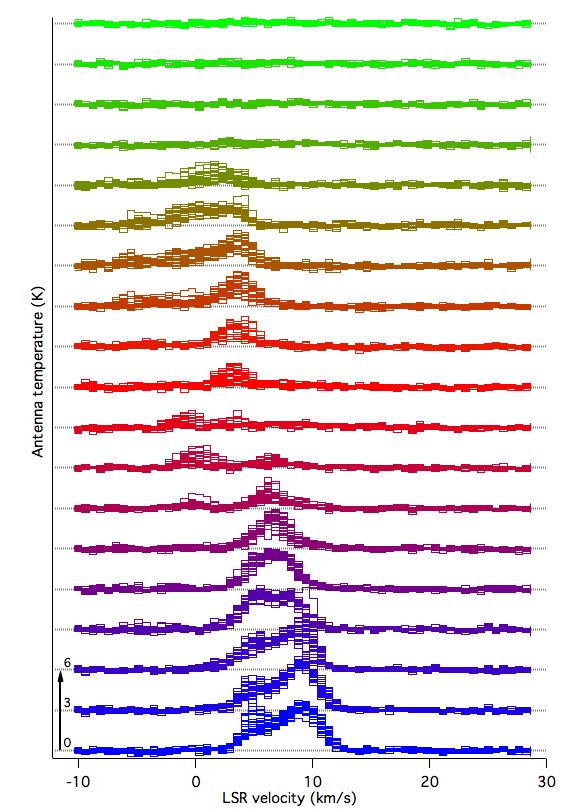}
\caption{New $^{12}$CO measurements: shown are spectra interpolated within the spectral cube for each of the 29 directions distributed in each of the 19 XMM fields. The spectra are displaced by 3 K from one XMM field to the other. The isolated shadowing molecular cloud +9$\fdeg$ is characterized by a smaller radial velocity than low latitude clouds.}
\label{New_CO_PROFILES}
\end{figure}

\begin{figure*}
\centering
\includegraphics[width=0.85\linewidth, angle=0]{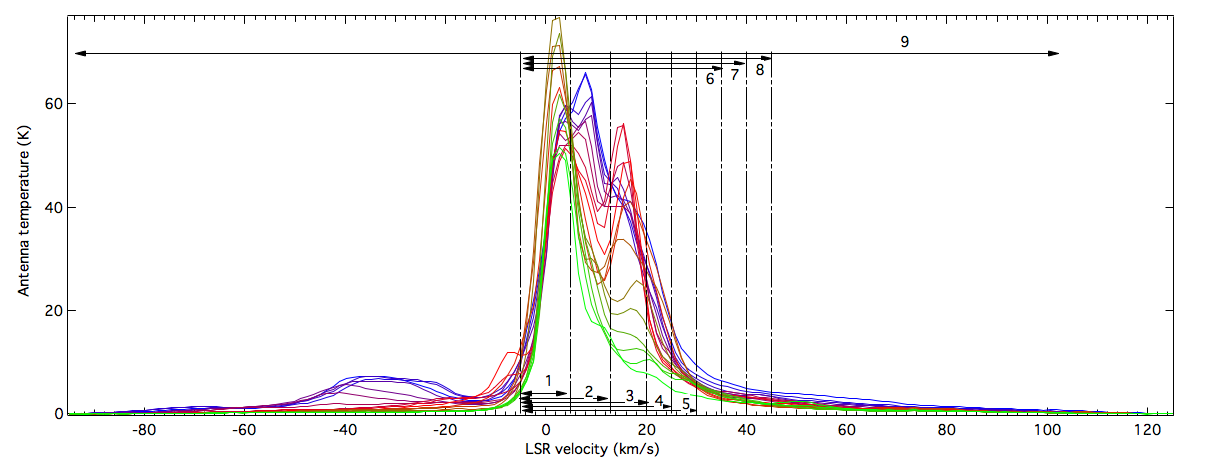}
\caption{Averaged HI 21cm EBHIS spectra for the 19 fields. Vertical lines indicate the limits of the velocity intervals used in section 4.}
\label{EBHIS_SPECTRA} 
\end{figure*}

\subsection{Gas}

\subsubsection{New CO Survey data}

Since the region of the XMM fields was only sampled every $1/4$ degree by the whole Galaxy CO survey of \cite{Dame01}, we undertook new more finely spaced CO observations of the region with the Center for Astrophysics 1.2 m telescope, the same instrument used for the Dame et al. survey.   The region l = 26.75 -  30.75, b = 4 – 12 was sampled slightly better than every beam width (every 7.5’) with a 8.4’ beam.   A total of 1536 positions were observed with integration times of $\simeq$1 min, but automatically adjusted based on the instantaneous system temperature to obtain a uniform  survey sensitivity of 0.2 K in 0.65 km s$^{-1}$ spectra channels.    Since existing data in the region revealed that all of the emission was confined to the velocity range -5 to +25 km s$^{-1}$, we were able to frequency switch by 15 MHz while keeping all of the emission within the 64 MHz band of the spectrometer.  The spectra were subsequently folded and 5th order polynomial baselines were removed. During the observations, the telluric emission line of CO appeared at velocities between -45 and -39 km s$^{-1}$, well displaced from the celestial emission, and could be easily removed from the data. 

Fig. \ref{New_CO_MAP} shows the velocity-integrated CO brightness in the XMM measurements area. Apart from the latitudinal decrease there is an isolated cloud at $\simeq$ +9$\fdeg$ latitude, with a marked correlation between CO emission and X-ray absorption in this region, which demonstrates that the NPS source is beyond this cloud. As shown from the individual spectra displayed in Fig \ref{New_CO_PROFILES}, for the entire set of sightline the CO velocity is comprised between -5 and +13 km s$^{-1}$, and the high latitude cloud has a LSR velocity between -5  and +2 km s$^{-1}$, significantly smaller than lower latitude molecular material.

\subsubsection{Use of combined HI and CO data}

Here we are comparing the observed N$_{Habs}$ to total columns N(HI)+ 2N(H2) derived from HI 21cm and CO spectra. We have used the new $^{12}$CO survey data described above and the recently published HI 21cm EBHIS survey data \citep{Winkel16} in a combined study of the IS gas in the directions of the XMM fields. EBHIS and CO spectra were interpolated for the same 29 directions within each XMM field, taking benefit of the high resolution of both datasets, then FOV averaged.  Fig.\ref{EBHIS_SPECTRA} displays the resulting average EBHIS HI spectra which, as expected, reveal velocity components at smaller and higher LSR velocities compared to CO. Based on the main components that appear in the spectra and appearing in the figure, we considered a series of LSR velocities intervals: (1) -5 to +5 km s$^{-1}$, i.e. nearby gas corresponding to the first and most intense component, that includes the +9$\fdeg$ molecular cloud, (2)  -5 to +13 km s$^{-1}$, i.e. the first two components and the totality of the detected CO, (3) -5 to +20 km s$^{-1}$, i.e. the extension to the third strong component seen at all latitudes at velocities around +15 km s$^{-1}$, (4) -5 to +25 km s$^{-1}$, including the apparent extension at high latitude of the previous component. We also consider the following extensions: (5) -5 to +30 km s$^{-1}$, (6) -5 to + 35 km s$^{-1}$, (7) -5 to + 40 km s$^{-1}$, (8) -5 to + 45 km s$^{-1}$, and finally (9)  -100 to +100 km s$^{-1}$. For each velocity interval we can integrate the HI and CO profiles and compute the corresponding total gas profiles N(H) along the 19 fields, for a given X$_{CO}$=N(H$_{2}$)/W(CO). 

\begin{figure}
\centering
\includegraphics[width=0.98\linewidth, angle=0]{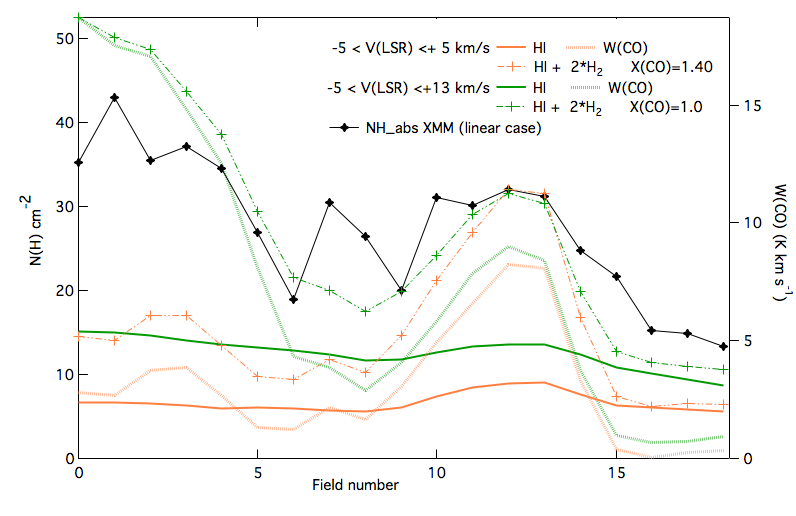}
\caption{Comparison between NPS X-ray absorbing gas column for the 19 fields (black signs) and columns from radio emission data: shown are field-of-view-averaged (H) columns deduced from HI 21 cm EBHIS data (solid lines) and from new CO data (dotted lines) for the first two velocity intervals (yellow and green respectively). Also plotted are the total N(H) from HI and CO assuming the maximal X$_{CO}$ conversion factor compatible with the X-ray absorption and the b=+9$\fdeg$ molecular cloud shadow (see text).}
\label{NH_WCO1} 
\end{figure}

 Fig  \ref{NH_WCO1} shows both W(CO), N(HI) and N(HI)+2xN(H$_{2}$) for the first two velocity intervals. There is a well defined emission maximum in fields 10 to 14 which corresponds to the shadowing molecular cloud at b=9$\fdeg$ we previously discussed, whose LSR velocity is around -2 km s$^{-1}$, i.e. within this first interval.  Its velocity that is well defined thanks to the CO data is slightly lower than the one of the bulk of the Aquila Rift clouds and lower latitude clouds, but it seems geometrically associated to these lower latitude structures. Thus both the geometry and its low velocity suggest that it is a nearby structure linked to the main Aquila Rift clouds.  Given its small distance, and subsequently its small size, it is extremely likely that the NPS source, a very wide structure, is located beyond the totality of this cloud, providing a first firm upper limit on the X$_{CO}$ factor for this cloud. Fig \ref{NH_WCO1} shows the sum of N(HI) and the converted W(CO) for the first velocity interval -5,+5 km s$^{-1}$ which at high latitude is dominated by the emission from this cloud, here for X$_{CO}$= 1.4 x 10$^{20}$ cm$^{-2}$ K$^{-1}$ km$^{-1}$ s. With such a value, N(H)=N(HI)+2N(H2) reaches the X-ray absorbing H column, implying that X$_{CO}$=1.4 x 10$^{20}$ cm$^{-2}$ K$^{-1}$ km$^{-1}$ s is a strict upper limit to the X$_{CO}$ factor for this cloud. On the other hand, it is clear from the N(H) profiles in the figure that matching the absorption at all latitudes is totally precluded if we restrict the absorbing matter to this velocity interval, and that addition of gas at higher velocity is required. This is in agreement with the absorption velocities found in optical spectra of stars within 500 pc, showing heliocentric absorptions comprised between -20 and and -5 km s$^{-1}$, i.e. corresponding to LSR velocities in both the first and second intervals (V(LSR)-V(Helio)$\simeq$ 15-17 km s$^{-1}$). Using the second, wider interval (-5, +13 km$^{-1}$), it is possible to find a much better match of the XMM absorptions, but this time 
X$_{CO}$ is more strongly limited and constrained below $\simeq$1.0 x 10$^{20}$ cm$^{-2}$ K$^{-1}$ km$^{-1}$ s, as shown in the figure. This value is much lower than the Galaxy-averaged value of $\simeq$ 1.8 x 10$^{20}$ cm$^{-2}$ K$^{-1}$ km$^{-1}$ s  \citep{Dame01}, but is on the order of the factors found from combined Fermi-Planck analyses of nearby clouds (see, e.g., \cite{PlanckFermi15} or \cite{Remy15}). 

\begin{figure}
\centering
\includegraphics[width=0.98\linewidth, angle=0]{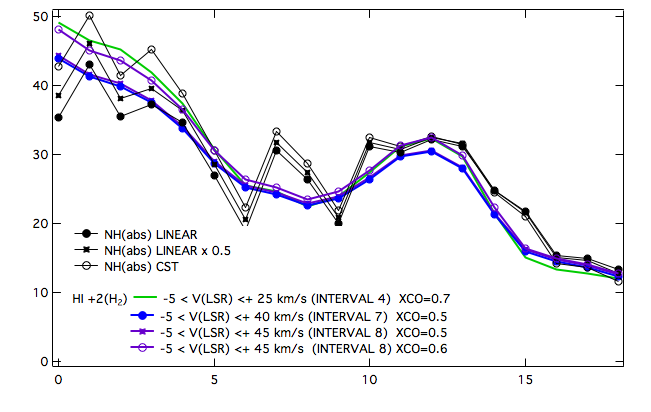}
\caption{Comparison between NPS X-ray absorbing gas column for the 19 fields (black diamonds) and total columns from HI and CO for wider velocity intervals. Solutions that best match the three fitted NH(abs) cases are indicated by the same markers than the corresponding XMM profiles.}
\label{NH_WCO2} 
\end{figure}

Fig. \ref{NH_WCO2} shows the extension of the N(H) computations to more extended velocity intervals. We have also included in Fig. \ref{NH_WCO2} the three fitted columns found for the constant NPS flux case, the linearly varying flux with maximum slope, and the intermediate case with a reduced slope (see section 2).
It is clear from both figures  \ref{NH_WCO1} and \ref{NH_WCO2} that the global N$_{Habs}$ profiles can be well reproduced for several velocity intervals and X$_{CO}$ factors, but that none of the model profiles deduced from HI and CO reproduces perfectly well their field-to-field variations: for a few directions (especially fields 6 and 9) there are discrepant results that are impossible to account for by varying solely the velocity intervals or X$_{CO}$. This variability deserves further studies that are beyond the scope of the present global analysis.
 
For each of the three hypotheses used to derive the N$_{Habs}$ profiles and for various values of the model parameters describing N(HI)+ 2N(H$_{2}$), a residual (sum over the 19 observing points of (data-model)$^{2}$) can be computed. We first searched for the minimum residual between N(H)$_{abs}$ and N(HI)+2N(H$_{2}$) for the three cases, varying X$_{CO}$ by steps of 0.1 x 10$^{20}$ cm$^{-2}$ K$^{-1}$ km$^{-1}$ s and considering all velocity intervals.  Fig. \ref{NH_WCO2} shows the corresponding total columns of H nuclei based on EBHIS and the new CO survey for the three optimal solutions, using unweighted data-model deviations (see below). In the three cases X$_{CO}$ is found to be very low, on the order of 0.5-0.6 x 10$^{20}$ cm$^{-2}$ K$^{-1}$ km$^{-1}$ s, and velocity intervals with an upper limit beyond +35 km s$^{-1}$ are required. We additionally displayed in Fig. \ref{NH_WCO2} a profile corresponding to the -5, +25 km s$^{-1}$ velocity interval (4) and X$_{CO}$=0.7 x 10$^{20}$ cm$^{-2}$ K$^{-1}$ km$^{-1}$ s. It is clear from the figure that such a solution, although not favored by minimization of residuals, is also quite close to the N(H)$_{abs}$ profile for the constant flux case. As a matter of fact, from velocity intervals 3 to 8 there is little change in the HI columns, as can be seen from fig \ref{EBHIS_SPECTRA}, and no variation at all of H2 as can be derived from the CO spectra. As a consequence, all profiles are very similar, and it is difficult to preclude a solution, especially given the discrepancies we previously noticed for several directions. 

\begin{figure}
\centering
\includegraphics[width=0.98\linewidth, angle=0]{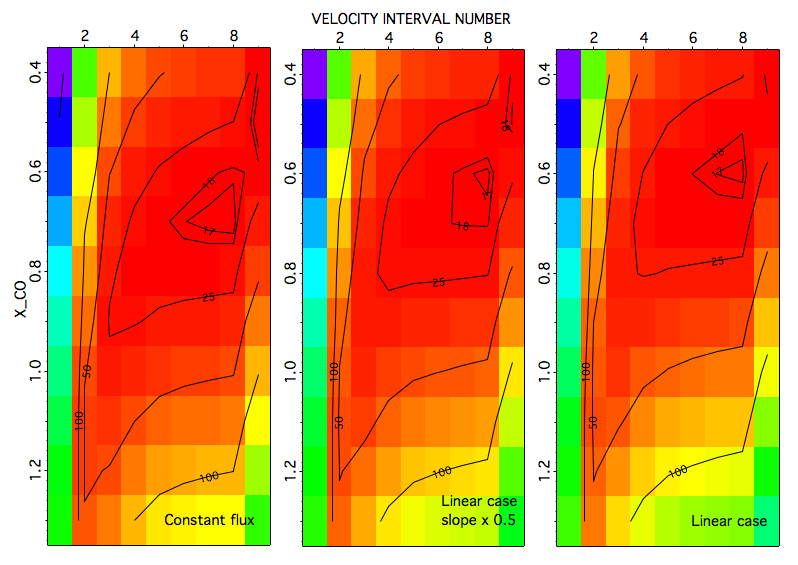}
\caption{Weighted residuals between the measured absorbing columns and the HI+H$_{2}$ columns in the three cases, as a function of X$_{CO}$ and the IS velocity interval.}
\label{residuals} 
\end{figure}

Despite those ambiguities, it is informative to perform a formal study of the uncertainties on the parameters X$_{CO}$ and the velocity range. As was said before, the values of  N$_{Habs}$ are determined for various hypotheses about the variation of the X ray emission with latitude: constant, linear variations with two different slopes. For a given hypothesis, the formal error bar on N$_{Habs}$ derived form the X-ray analysis on is very small, on the order of 1. to 1.5 x 10$^{20}$ cm$^{-2}$ as can be seen in Tables \ref{tabfields} and \ref{tabfields2}, because the number of data points in the spectra is very large. However, given the uncertainty on the hypothesis on the flux latitudinal  variation, there is a systematic error bar on N$_{Habs}$ whose order of magnitude can be estimated by the range of N$_{Habs}$ values found for the various hypotheses, i.e. between 0.2 and 8 x 10$^{20}$ cm$^{-2}$ depending on the direction (see Fig. \ref{NH_WCO2}). Accordingly, we estimated the total uncertainty for each direction as the quadratic sum of the statistical error bar for each fit solution and the difference between the two extreme N$_{Habs}$ solutions (namely the constant case and the linear case), scaled by a coefficient $\alpha$ on the order of one. A classical way to estimate the actual total uncertainty it is to find the magnitude of error bar that provides a sum of weighted residuals equal to the number of data points minus the number of free parameters, the velocity interval number and X$_{CO}$, i.e. here 19-2=17. We thus adjusted $\alpha$ in such a way this condition is fulfilled, which gives $\alpha$= 1.35, 1.40 and 1.47 for the three cases respectively. The corresponding sum of weighted residuals for the nine velocity intervals and X$_{CO}$ between 0.4 and 1.3 is shown in Fig \ref{residuals}. For the three cases shown in Fig \ref{residuals}, the residuals are minimum for velocity intervals 7 and 8, i.e. for the widest velocity intervals considered, except the last one (-100,+100 km s$^{-1}$). In the case of random, Gaussian errors, the 1 sigma domain of model parameters X$_{CO}$  and velocity range numbers would be delimited in Fig. \ref{residuals} by the iso-contour $\chi^{2}$=18 ($\delta \chi^{2}$=1). Such a condition is not fulfilled here, as already discussed, nevertheless we can draw some conclusions from the global pattern. The figure shows that three velocity intervals are definitely excluded, intervals 1, 2 and 9, and that interval 3 is also significantly disfavored, but that apart from these intervals the $\chi^{2}$ is varying very weakly, a consequence of the very little amount of atomic gas between +25 and +45 km s$^{-1}$. According to Galactic rotation models (see, e.g., \cite{Vallee08}) the LSR radial velocity of Sagittarius Arm gas is $\simeq$+20, +30 km s$^{-1}$, which implies that the NPS source may be located beyond the Arm. Fig. \ref{EBHIS_SPECTRA} and Fig. \ref{residuals} show that the velocity interval 9 is precluded essentially because of the significant contribution of gas at large negative velocities ($\leq$ -20 km s$^{-1}$). Galactic rotation models locate such gas at very large distance, well beyond the Galactic Centre (GC). This suggests that the NPS is, as expected, closer than the GC. On the other hand, the amount of gas between +45 and +100 km s$^{-1}$ is also very small, which means that it is difficult to preclude a contribution from far beyond Sagittarius, and an even more distant NPS. 

It is also clear from Fig \ref{residuals} that values of X$_{CO}$ higher than 0.75 x 10$^{20}$ cm$^{-2}$ K$^{-1}$ km$^{-1}$ s are very unlikely, as the residuals jump very abruptly beyond this value. The highest value of $\simeq$ 0.75 x 10$^{20}$ cm$^{-2}$ K$^{-1}$ km$^{-1}$ s is found for the constant flux case, while for the two cases that are favoured by X-ray spectral fitting the upper limit is lower, on the order of 0.60 x 10$^{20}$ cm$^{-2}$ K$^{-1}$ km$^{-1}$ s. This confirms the low CO-H$_{2}$ conversion factor associated with the nearby molecular clouds inferred from Fermi-Planck analyses.

\section{Conclusions and discussion}

The southern terminus of the X-ray NPS has been analyzed with unprecedented detail and spatial resolution by means of a dedicated series of XMM-Newton measurements. Spectral fitting to the whole dataset provided the X-ray absorbing columns of IS gas in 19 directions spanning a Galactic latitude interval of 5.5 degrees at l$\simeq29$$\fdeg$. The absorbing columns vary from 1.3 to 4.3 x 10$^{21}$ cm$^{-2}$ between b=+11$\fdeg$.1 and b=+5$\fdeg$.6. A shadowing cloud visible in CO maps is clearly detected at b= $\simeq$+9$\fdeg$. Such measurements allowed for the first time a detailed comparison with tracers of the IS dust gas and gas distribution, both distance-limited data and total LOS emission-based data, in particular a dedicated radio $^{12}$CO high resolution survey and ground-based absorption measurements in support to the XMM program. We have obtained the following results:

1) There is compelling evidence from surface brightnesses, hardness ratios and absorbing columns that the southern terminus of the North Spur is fully absorption bounded and consequently that the source extends farther toward the Plane. 

2) For the three stars located in the XMM fields, estimated IS gas columns N$_{Habs}$ deduced from absorption lines are below the X-ray absorbing columns, implying a first minimum distance to the NPS inner boundary of 260 pc. 

3) The set of target stars was used to validate local dust maps  in the NPS area and to estimate the integrated reddening along the XMM directions up to 300 pc. Reddenings are converted into gas columns that are compared with N$_{Habs}$ for the 19 XMM fields. This comparison shows that the distance to the X-ray source front is definitely beyond 300 pc.

4) Independently, the comparison with the larger scale Pan-STARRS (PS) 3D dust maps \citep{Green15} also implies a minimal distance to the NPS of 300 pc, in agreement with evidence from recent studies based on other X-ray data, 3D tomography of dust or radio polarization \citep{Sofue15,Puspitarini14,Wolleben07}. The comparison favours a larger distance,  and uncertainties on the calibrations and the conversion factors allow for as much as 4 kpc. As a matter of fact, the low amount of reddening between 600 pc and 4 kpc (see Fig \ref{LOS_ALL_NH}) precludes a precise upper limit. On the other hand, a large distance is independently favoured by the intercomparison of the correlations between  N$_{Habs}$ and the PS color excesses at various distances. As a matter of fact, the correlation improves by increasing the sightline length from 400 pc to 4 kpc. Part of the effect may be linked to the improved statistics of the PS measure, however, and more work is needed to disentangle the two effects. 

5) The absorbing column N$_{Habs}$ latitude profile deduced from the X-ray spectral fitting is found to correlate more tightly with the reddening deduced from Planck dust optical depths than with any distance-limited reddening, suggesting that the emission originates beyond a very large fraction of the matter the Planck emission is tracing, i.e. potentially several kpc away. In terms of absolute values, the correspondence between the X-ray absorbing columns and Planck $\tau_{353GHz}$-based reddenings  results in N(H)/E(B-V) = 4.1 cm$^{-2}$ mag$^{-1}$,  a value that matches well recent determinations from Fermi and Planck data joint analyses \citep{PlanckFermi15}. 

6) The existence of a shadowing molecular cloud at b=+9$\fdeg$ allows the use of the X-ray absorbing columns and emission data to constrain the CO-H$_{2}$ conversion factor X$_{CO}$ below 1.0 x 10$^{20}$ cm$^{-2}$ K$^{-1}$ km$^{-1}$ s.

7) The combination of X-ray absorbing columns and emission data  for the entire XMM path constrains the average X$_{CO}$ below 0.75 x 10$^{20}$ cm$^{-2}$ K$^{-1}$ km$^{-1}$ s, with the most probable value as low as 0.6 10$^{20}$ cm$^{-2}$ K$^{-1}$ km$^{-1}$ s. Such values are unusually low, but X$_{CO}$ factors comprised between 0.6 and 1.1 10$^{20}$ cm$^{-2}$ K$^{-1}$ km$^{-1}$ s have been recently derived from combined Fermi-Planck studies of local clouds \citep{PlanckFermi15, Remy15}. 

 8) Absolute values of X-ray absorbing columns and their latitude profiles are compatible with HI and CO based columns of gas at  -5 $\leq$ V$_{LSR}$ $\leq$ +25 to +45 km s$^{-1}$, with the broadest interval being favoured. The large uncertainty on the velocity interval of the gas absorbing the NPS is due to the very limited amount of gas at positive velocities between +25 and +45 km s$^{-1}$. According to kinematical models of Galactic rotation (e.g. \cite{Vallee08}), at l=+29$\fdeg$ the radial velocity of the Sagittarius Arm gas is on the order of +20,+30  km s$^{-1}$), resulting in a potential location of the NPS in front of or beyond this Arm depending on the actual Sagittarius velocity range, with the latter case being more likely. Because there is a very small additional amount of gas faster than +45 km s$^{-1}$, a larger (positive) interval and a subsequent even more distant NPS can not be precluded. However, as discussed in section 4, more work is needed to understand the observed discrepancies and reduce the systematic uncertainties on the absorption profiles before a precise distance can be determined.

 The minimum distance to the NPS derived from this study definitely demonstrates the absence of link with the nearby Sco-Cen star-forming region. The high probability of a much larger NPS distance that comes out from comparisons with dust and gas absorption and emission raises one more time the question of a possible link between the Spur and outflows from the inner Galaxy (Fermi bubbles, Galactic wind).  \cite{PlanckXXV15} disfavor such a link based on the identification of northern and southern polarized emission structures with Loop I secondary arcs and the following geometrical arguments: - the strong North-South asymmetry of the NPS, - the absence of a pinched structure symmetric about the Galactic plane, - the absence of a trace of any interaction between NPS/Loop I and the Fermi bubbles. However, it can be argued that NPS-Loop I and the Fermi bubbles may trace completely distinct episodes of nuclear activity, in which case geometrical arguments become weaker. 
 
Refined analyses of the XMM spectra should help constraining further the NPS characteristics. Although absorbing columns are not expected to vary significantly after relaxation of the constraints on the component parameters that have been imposed here, allowance for departures from thermal emission, for non-solar ion abundances and abundance ratios, as well as the inclusion of a dedicated modeling of the heliospheric charge-exchange emission should all together put stronger constraints on the emission mechanisms.  This is beyond the scope of the present work, mainly devoted to comparisons with existing IS data. 
From the ISM distribution point of view, hopefully future 3D mapping of Galactic clouds and their distance and velocity assignments should allow to take larger benefit from the present study and better constrain the NPS source location. In particular, future measurements with Gaia should better constrain the cloud distribution from both parallax data and improved reddening estimates. In particular, accurate measurements of Sagittarius gas velocities should allow to use our results in a more definitive way. Thanks to Gaia too, young star associations in the Inner Galaxy that may give rise to a giant super-bubble and a structure similar to NPS-Loop I should also be identified and located. On the longer term, high sensitivity, spectral and spatial resolution Athena X-ray spectra \citep{Athena13} are expected to shed additional light on the NPS spectral characteristics and the nature of its source. 

%However, as can be seen from Pan-STARRS reddenings in Fig \ref{LOS_ALL}, most of the matter along the XMM sightlines is located within the first kiloparsec, with only a 5 to 30\% reddening increase from 1 to 6 kpc depending on directions. As a consequence, given the uncertainties in the X-ray absorbing columns and in the various conversion factors and calibrations we used, it is not possible to demonstrate that the NPS-Loop I source definitely lies much farther away than 300 pc. Still, our results 

%Hopefully the relaunched debate on the NPS-Loop I location will be definitely settled in a near future.
%Previous bursting nuclear activity may have launched wind in a more empty halo
%Today's burst different in the sense that it expands in a previously filled halo

%To discuss: XCO and NH
%absorptions and emission radial velocities
%No one of the NARVAL stars shows velocities corresponding to VLSR=+15 
%nor VV Ser in the Serpens dark cloud at 415pc (voir le spectre ESPADONS KI magnifique)

\begin{acknowledgements}
K.K. acknowledges support from NASA grant NNX15AG24G. R.L. and I.G. acknowledge support from the French
National Research Agency (ANR) through the STILISM project. R.L. acknowledges telescope time funding by the CNRS program (Programme National de Physique et Chimie du Milieu Interstellaire) and thanks the TBL-Narval team at Pic du Midi for their efficient support and excellent service observing.
\end{acknowledgements}

\bibliographystyle{aa} % style aa.bst
\bibliography{mybib} % your references Yourfile.bib

\clearpage
\section{Appendix: interstellar KI absorption lines in nearby star spectra}

\begin{figure}[h!]
\centering
\includegraphics[width=0.98\linewidth, angle=0]{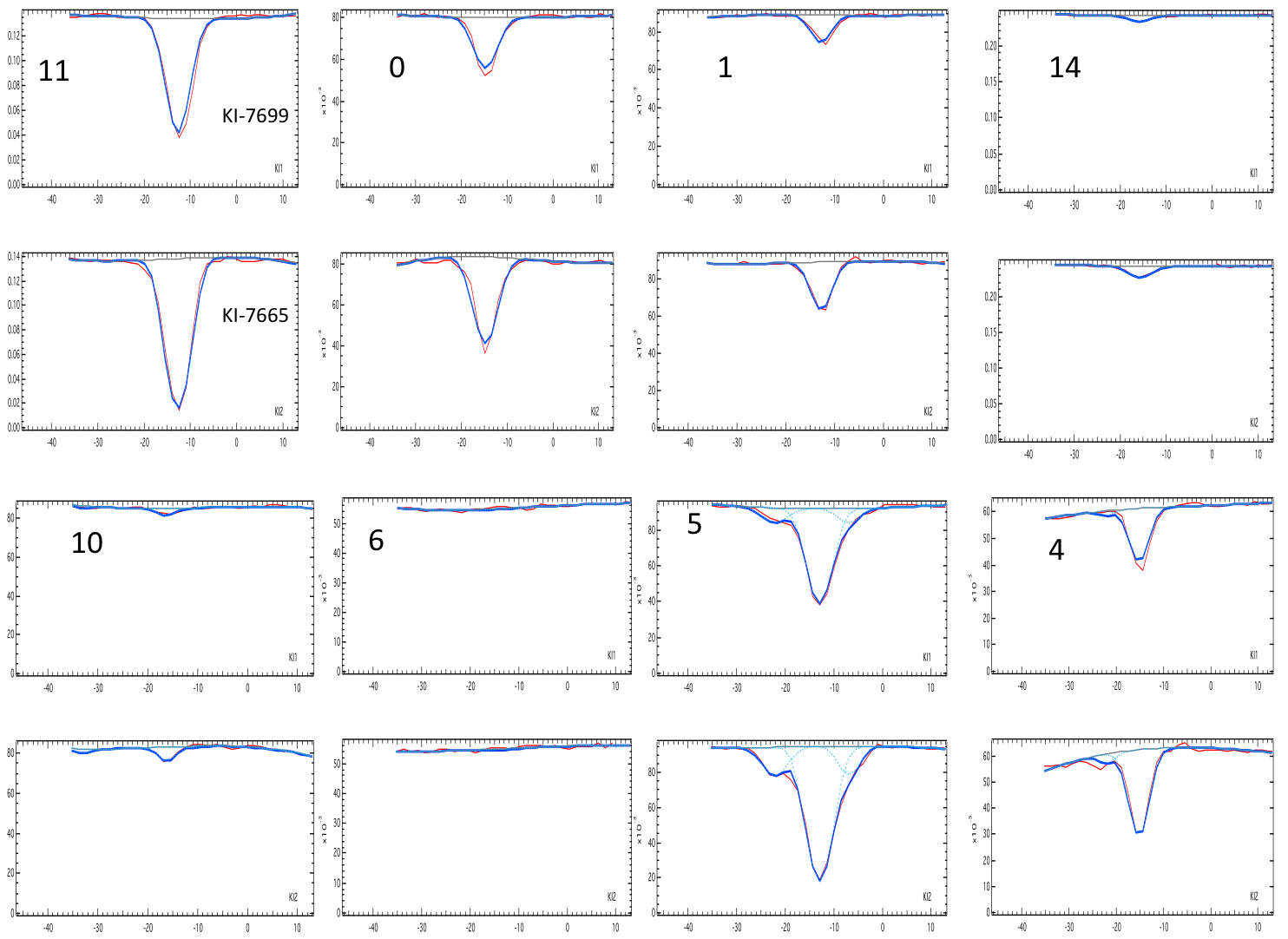}
\caption{Determination of the interstellar KI columns along the line-of-sight to the nearby stars of Table \ref{tabstars}. Numbers follow the convention of the table. Velocities are heliocentric. Flux units are arbitrary. For the eight early-type stars shown in the figure we performed a simultaneous fit of the two KI transitions.}
\label{starfit1} 
\end{figure}

\begin{figure}[h!]
\centering
\includegraphics[width=0.98\linewidth, angle=0]{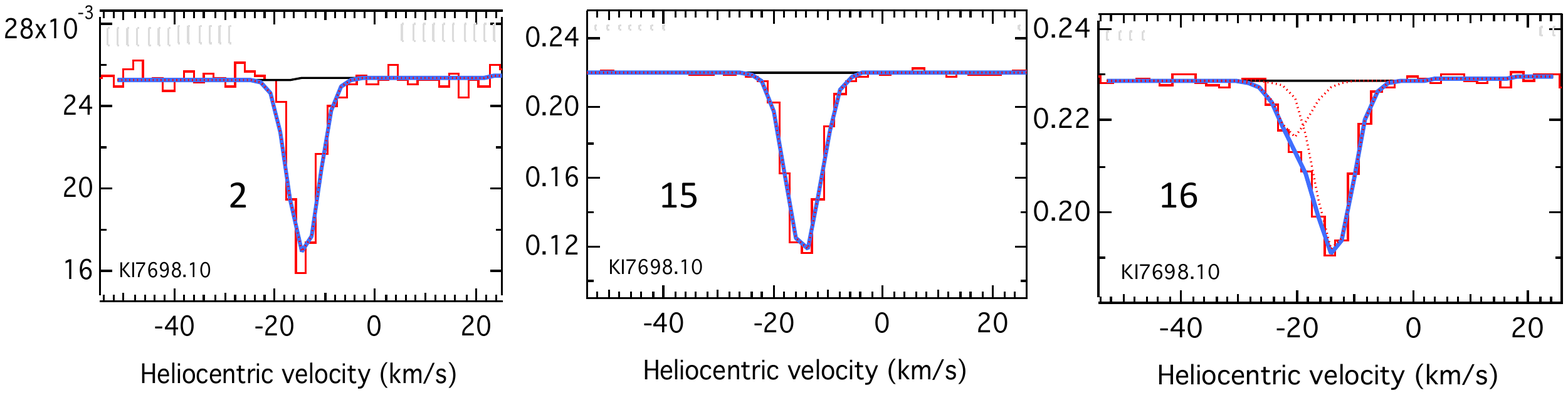}
\caption{Same as Figure \ref{starfit1} for three early-type stars strongly contaminated by telluric absorption. KI was determined from KI 7698\AA\ only.}
\label{starfit2} 
\end{figure}

\begin{figure}[h!]
\centering
\includegraphics[width=0.32\linewidth, angle=0]{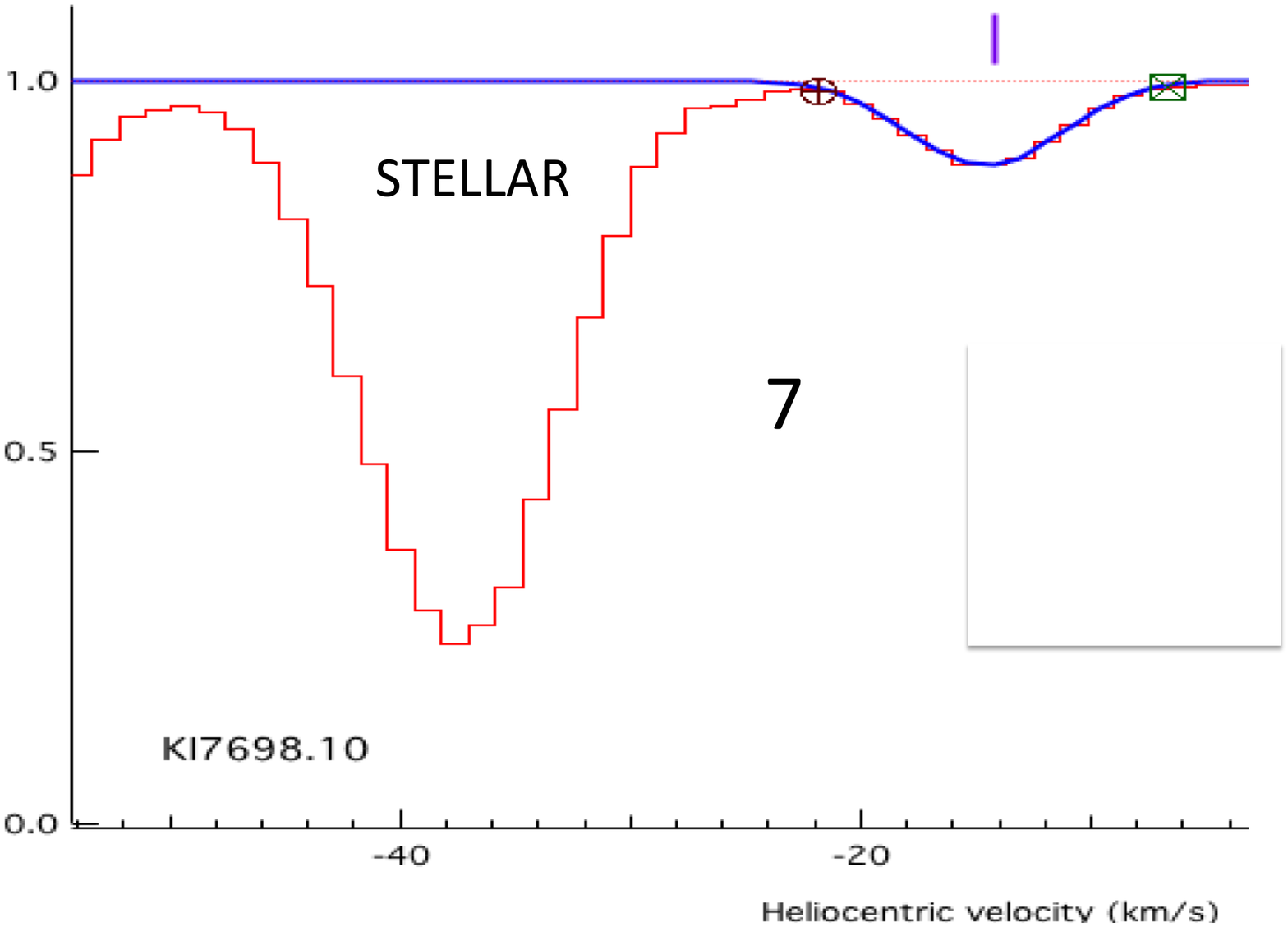}
\includegraphics[width=0.32\linewidth, angle=0]{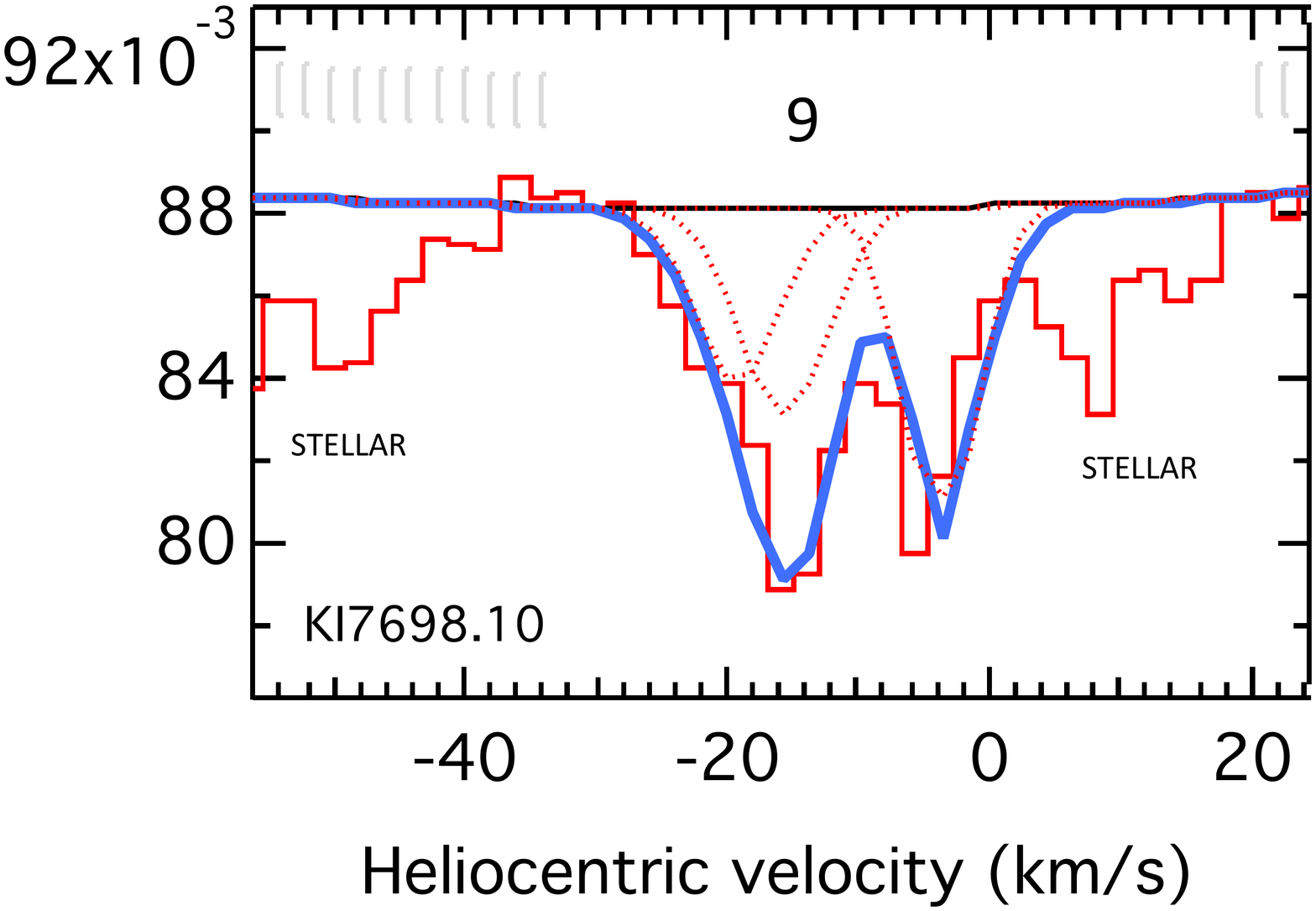}
\includegraphics[width=0.32\linewidth, angle=0]{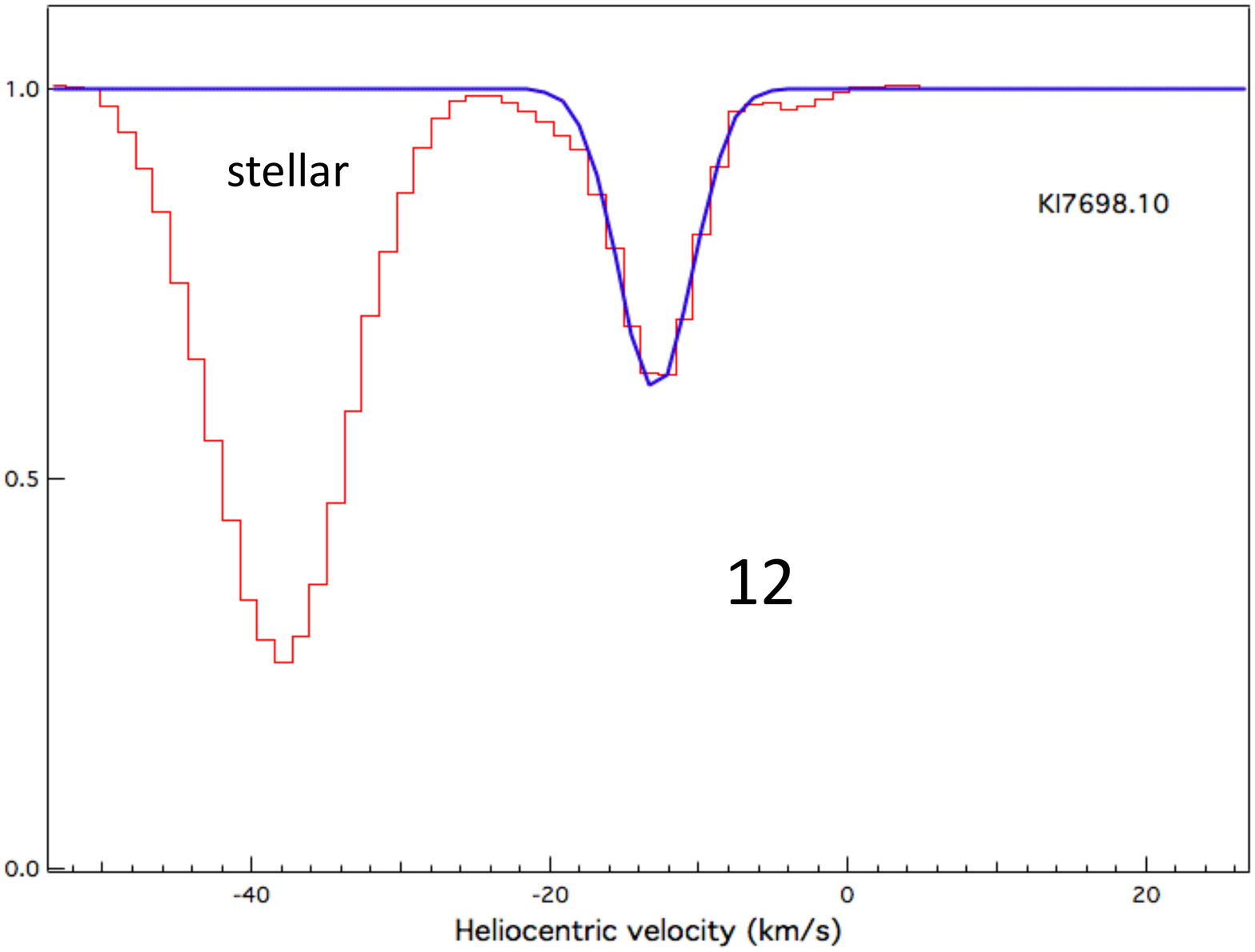}
\caption{Same as Figure \ref{starfit1} for three late-type stars with a large velocity shift between the stellar KI line and the IS KI lines. KI was determined from KI 7698\AA\ only. For star 9 there is a red-shifted, partially overlapping and weak stellar line not included in the adjustment.}
\label{starfit3} 
\end{figure}

\begin{figure}[h!]
\centering
\includegraphics[width=0.32\linewidth, height=4.cm,angle=0]{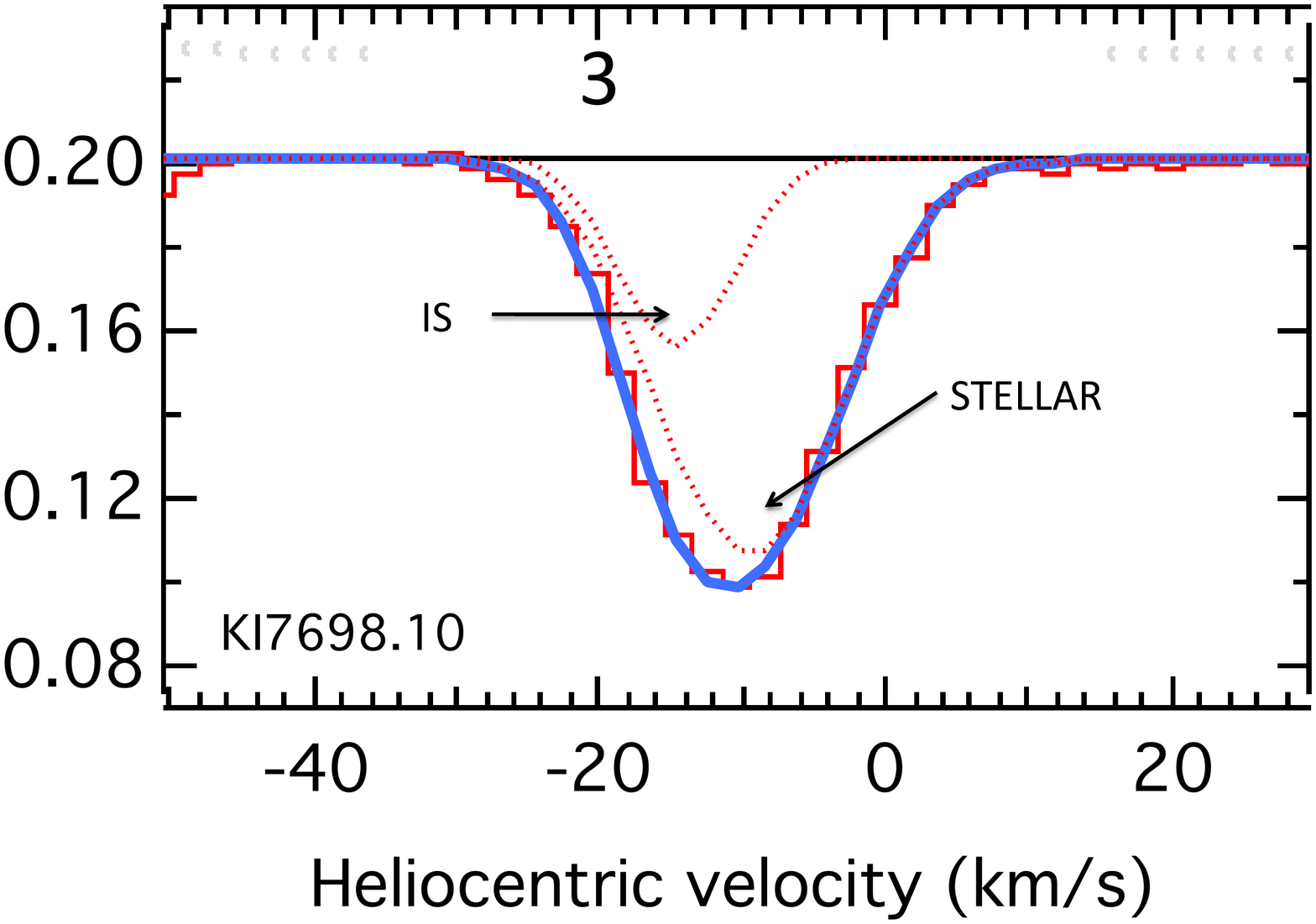}
\includegraphics[width=0.32\linewidth, height=3.5cm,angle=0]{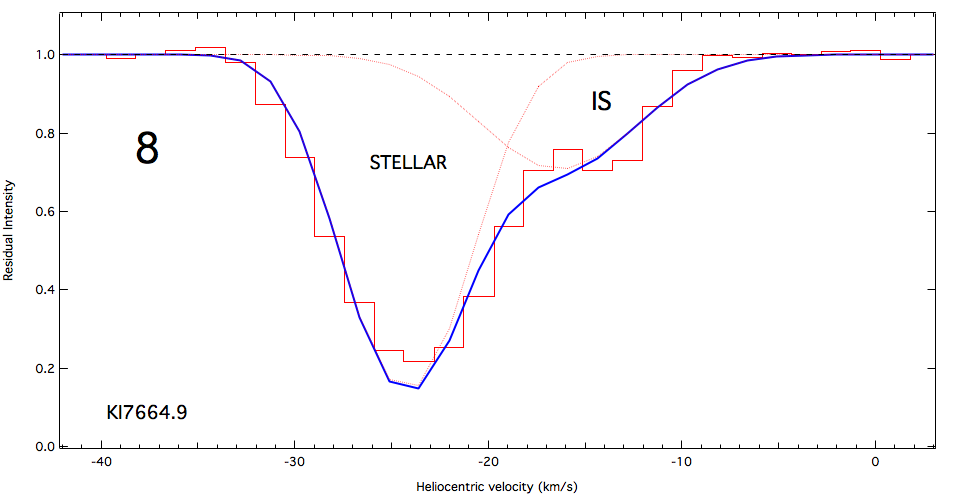}
\includegraphics[width=0.32\linewidth, height=3.5cm,angle=0]{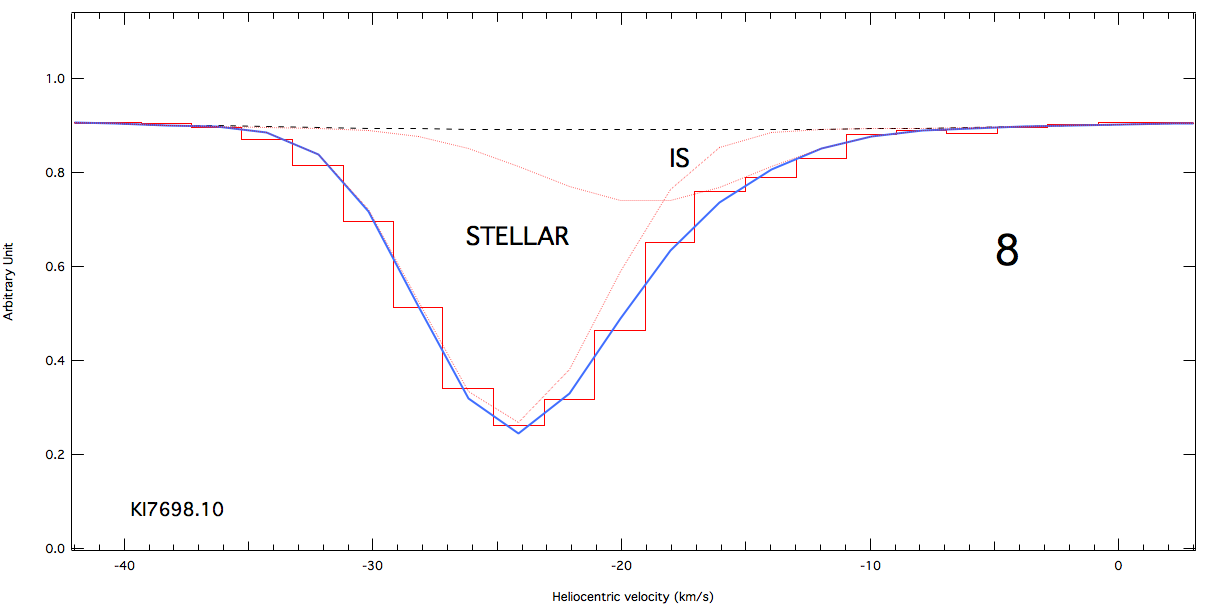}
\caption{Same as Figure \ref{starfit1} for two late-type stars with overlapping stellar and IS KI lines. The stellar line shift has been preliminarily determined from the strong sodium lines and imposed. The stellar KI line is fitted as an artificial IS line with a high temperature. KI was determined from the 7698\AA\ line only for star 3, and from the two transitions for star 8 (the two adjustments are shown in separate windows).}
\label{starfit4} 
\end{figure}

Figures \ref{starfit1}, \ref{starfit2},  \ref{starfit3} and \ref{starfit4} display the various model adjustments performed in order to extract IS KI absorbing columns from the TBL-NARVAL optical spectra of the stars listed in Table \ref{tabstars}. The fitting methods  vary according to the stellar type, the telluric contamination and the velocity shift between the stellar and IS KI lines. The resulting columns are listed in Table \ref{tabstars}.

\end{document}